\documentclass{JHEP3}
\usepackage{amsmath}
\usepackage[dvips]{graphicx}
\usepackage{epsf}

\renewcommand{\d}{\mathrm{d}}
\newcommand{\I}{\mathrm{i}}
\newcommand{\e}{\mathrm{e}}
\newcommand{\p}{\partial}

\newcommand{\cA}{\mathcal{A}}
\newcommand{\cB}{\mathcal{B}}
\newcommand{\cC}{\mathcal{C}}

\newcommand{\cG}{\mathcal{G}}
\newcommand{\cL}{\mathcal{L}}
\newcommand{\cK}{\mathcal{K}}

\newcommand{\cO}{\mathcal{O}}
\newcommand{\cR}{\mathcal{R}}
\newcommand{\cV}{\mathcal{V}}
\newcommand{\cW}{\mathcal{W}}

\newcommand{\de}{\delta}
\newcommand{\ep}{\varepsilon}
\newcommand{\zb}{\bar{z}}
\newcommand{\si}{\sigma}

\newcommand{\SU}{\mathrm{SU}}

\newcommand{\USp}{\mathrm{USp}}
\newcommand{\U}{\mathrm{U}}

\newcommand{\inst}{\mathrm{inst}}
\newcommand{\pert}{\mathrm{pert}}
\newcommand{\class}{\mathrm{class}}
\newcommand{\mn}{\text{min}}
\newcommand{\mx}{\text{max}}

\newcommand{\2}{\sqrt{2\,}}
\newcommand{\half}{\tfrac{1}{2}}
\newcommand{\quart}{\tfrac{1}{4}}

\newcommand{\tab}{\quad\,}
\newcommand{\lp}{\begin{pmatrix}}
\newcommand{\rp}{\end{pmatrix}}

\DeclareMathOperator{\re}{\mathrm{Re}}
\DeclareMathOperator{\im}{\mathrm{Im}}

\DeclareSymbolFont{AMSb}{U}{msb}{m}{n}
\DeclareMathSymbol{\fieldR}{\mathalpha}{AMSb}{"52}
\DeclareMathSymbol{\fieldZ}{\mathalpha}{AMSb}{"5A}

\preprint{ITP--UU--05/23, SPIN--05/18, FSU--TPI--04/05, hep-th/0506097}

\title{Membrane Instantons and de~Sitter Vacua}

\author{Marijn Davidse, Frank Saueressig \\
Institute for Theoretical Physics \emph{and} Spinoza Institute \\
Utrecht University, 3508 TD Utrecht, The Netherlands \\
E-mail: \email{M.Davidse@phys.uu.nl}, \email{F.S.Saueressig@phys.uu.nl}}

\author{Ulrich Theis \\
Institute for Theoretical Physics, Friedrich-Schiller-University
Jena, \\ Max-Wien-Platz 1, D-07743 Jena, Germany \\
E-mail: \email{Ulrich.Theis@uni-jena.de}}

\author{Stefan Vandoren \\
Institute for Theoretical Physics \emph{and} Spinoza Institute \\
Utrecht University, 3508 TD Utrecht, The Netherlands \\
E-mail: \email{S.Vandoren@phys.uu.nl}}

\abstract{%
We investigate membrane instanton effects in type IIA strings
compactified on rigid Calabi-Yau manifolds. These effects contribute
to the low-energy effective action of the universal hypermultiplet.
In the absence of additional fivebrane instantons, the quaternionic
geometry of this hypermultiplet is determined by solutions of the
three-dimensional Toda equation. We construct solutions describing
membrane instantons, and find perfect agreement with the string
theory prediction. In the context of flux compactifications we 
discuss how membrane instantons contribute to the scalar potential
and the stabilization of moduli. Finally, we demonstrate the existence
of meta-stable de~Sitter vacua.}

\keywords{Superstring Vacua, p-branes, Nonperturbative Effects}

\begin{document} 

\section{Introduction}

A central question in string theory is the existence and viability of
``semi-realistic'' four-dimensional ground states. In this context
studying the vacuum structure arising from flux compactifications has
recently attracted considerable attention. In particular, \cite{KKLT}
(KKLT) provided a qualitative picture for obtaining meta-stable
de~Sitter (dS) vacua from compactifications of the type IIB string, in
which fluxes and non-perturbative instanton effects play a crucial role.
In this paper we consider membrane instanton corrections arising in the
compactification of the type IIA string on rigid Calabi-Yau threefolds
(CY$_3$) and show that including background fluxes and these
non-perturbative corrections can provide another scenario to stabilize
all hypermultiplet moduli at a meta-stable de~Sitter vacuum.

The four-dimensional low-energy effective actions for string
compactifications preserving some supersymmetry are supergravity actions
coupled to matter multiplets. When fluxes are turned on, one typically
obtains gauged supergravities with a potential for the scalar fields of
the matter multiplets. The properties and extrema of such potentials are
of great importance for string cosmology, since they determine the
vacuum structure of the theory. In recent years, string theorists have
searched intensively for models in which the potential admits vacua with
a (small) positive cosmological constant. It turned out that it is
fairly difficult to realize such vacua in string theory, as they can
only be meta-stable (see e.g.\ \cite{S} for a review).

A qualitative picture on how such vacua can be obtained was given in
\cite{KKLT} in the context of type IIB flux compactifications on
orientifolds. In this case the four-dimensional effective action has
$N=1$ supersymmetry, and the potential is determined by a holomorphic
superpotential. The KKLT scenario relies on three contributions to the
superpotential: first there is a classical contribution coming from
fluxes which stabilizes all moduli except the volume modulus which does
not enter into a scalar potential of no-scale type. This modulus is then
stabilized by a non-perturbative contribution to the potential due to
D-instantons or gaugino condensation. These two ingredients stabilize
all moduli in a supersymmetric AdS vacuum. In the third step a (small)
positive energy contribution, as e.g.\ an anti-D3-brane, is added which
lifts the AdS vacuum to a positive cosmological constant. Since its
first proposal, possible realizations of this scenario either within
type IIB orientifold compactifications or their F-theory descriptions
have been studied intensively \cite{EGQ,BB2,BM2}.

One of the goals of this paper is to provide an alternative scenario in
the context of type IIA string theory compactified on a CY$_3$. Without
including background fluxes the LEEA arising from these
compactifications is a four-dimensional $N=2$ supergravity action
coupled to $h_{1,1}$ vector and $h_{1,2}+1$ hypermultiplets. There is no
scalar potential and the scalars (moduli) of the theory parameterize
flat directions. The coupling to $N=2$ supergravity requires the scalars
of the hypermultiplets to parameterize a quaternion-K\"ahler manifold
\cite{BW}. The dilaton that controls the quantum corrections sits in a
hypermultiplet (the universal hypermultiplet), and hence it is the
quaternionic geometry that receives quantum corrections. Besides
perturbative corrections, there are also non-perturbative instanton
effects obtained by wrapping Euclidean D-branes around supersymmetric
cycles of the internal manifold \cite{BBS,BB}. From the counting of
fermionic zero modes one can derive that they contribute to the
low-energy effective action. In the KKLT models they contribute to the
superpotential for the $N=1$ chiral multiplets, whereas in our case they
correct the hypermultiplet scalar metric.

In this paper, we focus on the special case of the universal
hypermultiplet, which can be obtained by compactifying on a rigid
CY$_3$, having $h_{1,2}=0$. We restrict ourselves to rigid CY manifolds,
because we will be able to explicitly determine the instanton
corrections in this special case only. The general situation when more
complex structure moduli are present is technically more difficult
because of the complicated nature of the quaternion-K\"ahler geometry.
We believe, however, that our main conclusion will still persist in this
case.

The classical quaternionic geometry of the universal hypermultiplet is
well-known \cite{CFG}, and recently the perturbative corrections were
found in \cite{AMTV}, see also \cite{ARV}. Non-perturbatively, there are
both membrane and NS fivebrane instanton corrections \cite{BBS}, but in
this paper we shall consider membrane instantons only.\footnote{For work
on fivebrane instantons, we refer the reader to \cite{DTV}. Additional
references on hypermultiplet moduli spaces and instantons are
\cite{GMV,OV,A,GS,TV1,DDVTV,BM}. Furthermore a program towards
formulating an instanton calculus based on $N=2$ supersymmetric actions
with Euclidean signature was started in \cite{CMMS1,CMMS2}.} In this
case the constraints from quaternionic geometry are captured by
solutions of the three-dimensional Toda equation. This fact was, to our
knowledge, first observed in \cite{K1} (see also \cite{K2,CIV}). One of
the main results of this paper is that we construct new solutions of the
Toda equation that correspond to membrane instanton expansions. We
have not uniquely fixed the solution, and at each order in the instanton
expansion, there is still an undetermined integration coefficient that
can in principle be computed in string theory. The solution of the Toda
equation then determines the quaternion-K\"ahler (QK) metric in the
ungauged supergravity effective action.\footnote{Membrane instantons
were also considered in \cite{K1,K2}, but our analysis below differs
since we do \emph{not} assume the existence of a rotational symmetry
between the RR scalars in the UHM scalar metric. In fact our analysis
will show that this isometry is broken.} As we will show, our results
are in complete agreement with the predictions made in \cite{BBS}.

Including background fluxes in the compactification leads to
four-dimensional $N=2$ gauged supergravity \cite{dWVP,DAFF,ABCDFFM}
where some isometries of the hypermultiplet scalar manifold are 
gauged \cite{LM,KKP}.\footnote{For an analysis on de Sitter vacua,
purely in the context of N=2 supergravity, we refer to \cite{FTVP}.}
This gauging induces a scalar potential in the LEEA which depends on the
geometrical quantities of the QK space, such as the moment maps and the
metric. It is therefore clear that the potential will receive quantum
corrections, determined e.g.\ by the QK metric. We must be careful with
this procedure, since isometries of the classical hypermultiplet moduli
space can be broken by quantum corrections. This is already the case
perturbatively \cite{AMTV,ARV}. Non-perturbatively, isometries can get
broken to discrete subgroups. To gauge an isometry in supergravity, the
standard methods require an unbroken and continuous isometry. However,
in the absence of fivebrane instantons, we explain how to find such an
isometry, and moreover we show how the corresponding potential can be
obtained from a flux compactification of the type discussed in
\cite{LM}.

In both the KKLT models with $N=1$, and as we will see, in our models
with $N=2$, it is crucial to take into account the quantum corrections
to the low-energy effective action. In particular, including the
instanton corrections to the potential is an essential step for
stabilizing the dilaton and finding meta-stable de~Sitter
vacua.\footnote{Based on the instanton corrected UHM of \cite{K2}, a
similar analysis, also indicating the existence of meta-stable dS vacua,
was performed in \cite{BM}.} This was the case in KKLT, and also applies
to our models.\footnote{Similar observations have also been made in
heterotic M-theory (see e.g.\ \cite{HetM}).} In our set-up, we only
study the hypermultiplet moduli in detail, and comment on the K\"ahler
moduli at the end of the paper. In that case, the potential only depends
on the hypermultiplet scalars and is determined by the solution of the
Toda equation. As our solution still contains undetermined integration
constants (which, in principle, should be determined by string theory),
it is therefore perhaps not too surprising that one can choose
coefficients that give de~Sitter vacua. In a way, choosing these
coefficients mimics stabilizing the volume modulus in the KKLT set-up.

The remainder of the paper is organized as follows. In section
\ref{sugra-des} we begin by describing the supergravity set-up for our
investigations. The moduli space metric of the universal hypermultiplet
is introduced and its possible quantum corrections are discussed
qualitatively. We then show in section \ref{PT} how this metric fits
into a general framework for four-dimensional QK geometries with one
isometry, which are governed by the three-dimensional Toda equation. In
section \ref{sect4} we derive the leading terms of a solution to this
equation describing non-perturbative quantum effects due to membrane
instantons. Section \ref{CY} is devoted to a comparison of our results
with string theory predictions on how these instanton corrections
contribute to the four-fermion coupling; we shall find perfect
agreement. Finally, in section \ref{dS} we investigate the effects of
these corrections on the scalar potential that arises by gauging the one
remaining isometry of the moduli space metric. It turns out that the
undetermined parameters can be such that the potential develops a local
meta-stable de~Sitter minimum. After the conclusions we provide
technical details in several appendices.

\section{Supergravity description} \label{sugra-des}

For type IIA string theories compactified on a CY$_3$ manifold, the
low-energy effective action is that of four-dimensional $N=2$
supergravity coupled to $h_{1,1}$ vector multiplets, $h_{1,2}$
hypermultiplets, and one tensor multiplet that contains the dilaton
\cite{BCF}. In the case of a rigid CY$_3$, there are no complex
structure moduli: $h_{1,2}=0$. Suppressing the vector multiplets, the
resulting four-dimensional low-energy effective action is that of a
tensor multiplet coupled to $N=2$ supergravity, and the bosonic part of
the Lagrangian at string tree-level is given by\footnote{Throughout this
paper, we work in units in which Newton's constant $\kappa^{-2}=2$.}
 \begin{align} \label{TM-action}
  e^{-1} \cL_\mathrm{T} & = - R - \frac{1}{2}\, \p^\mu \phi\, \p_\mu
    \phi + \frac{1}{2}\, \e^{2\phi} H^\mu H_\mu \notag \\*
 & \tab - \frac{1}{4}\, F^{\mu\nu} F_{\mu\nu} - \frac{1}{2}\, \e^{-\phi}
    \big( \p^\mu \chi\, \p_\mu \chi + \p^\mu \varphi\, \p_\mu
    \varphi \big) - \frac{1}{2}\, H^\mu \big( \chi \p_\mu \varphi -
    \varphi \p_\mu \chi \big)\ ,
 \end{align}
where $H^\mu=\tfrac{1}{6}\,\ep^{\mu\nu\rho\si}H_{\nu\rho\si}$ is the
dual NS 2-form field strength. The first line comes from the NS sector
in ten dimensions, and $\phi$ together with $H^\mu$ forms an $N=1$
tensor multiplet.
The second line descends from the RR sector. In particular, the
graviphoton with field strength $F_{\mu\nu}$ descends from the
ten-dimensional RR 1-form, and $\varphi$ and $\chi$ can be combined
into a complex scalar $C$ that descends from the holomorphic components
of the RR 3-form with (complex) indices along the holomorphic 3-form of
the CY$_3$. Notice the presence of constant shift symmetries on both
$\chi$ and $\varphi$. Together with a rotation on $\chi$ and $\varphi$
they form a three-dimensional subgroup of symmetries. 

The tensor multiplet Lagrangian \eqref{TM-action} is dual to the
universal hypermultiplet. This can be seen by dualizing the 2-form
into an axionic pseudoscalar field $\sigma$, after which one obtains
(modulo a surface term)
 \begin{align}\label{UH-action}
  e^{-1} \cL_\mathrm{UH} & = - R - \frac{1}{4}\, F^{\mu\nu} F_{\mu\nu}
    - \frac{1}{2}\, \p^\mu \phi\, \p_\mu \phi - \frac{1}{2}\,
    \e^{-\phi} \big( \p^\mu \chi\, \p_\mu \chi + \p^\mu \varphi\,
    \p_\mu \varphi \big) \notag \\*
 & \tab - \frac{1}{2}\, \e^{-2\phi} \big( \p_\mu \si + \chi \p_\mu
    \varphi \big)^2 \ .
 \end{align}
The four scalars define the classical universal hypermultiplet at
string tree-level, a non-linear sigma model with a quaternion-K\"ahler
target space $\SU(1,2)/\U(2)$ \cite{CFG}. The metric can be written as
 \begin{equation} \label{class-UHM}
  \d s^2 = G_{AB}\, \d\phi^A\, \d\phi^B = \d\phi^2 + \e^{-\phi} (\d
  \chi^2 + \d\varphi^2) + \e^{-2\phi} (\d\sigma +\chi \d\varphi)^2\ .
 \end{equation}
This manifold has an $\SU(1,2)$ group of isometries, with a
three-dimensional Heisenberg subalgebra that generates the following
shifts on the fields,
 \begin{equation} \label{Heis-alg}
  \phi \rightarrow \phi \ ,\qquad \chi \rightarrow \chi + \gamma\ ,
  \qquad \varphi \rightarrow \varphi + \beta\ ,\qquad \sigma
  \rightarrow \sigma - \alpha - \gamma\, \varphi\ ,
 \end{equation}
where $\alpha$, $\beta$, $\gamma$ are real (finite) parameters. 

Quantum corrections, both perturbative and non-perturbative, will break
some of the isometries and alter the classical moduli space of the
universal hypermultiplet, while keeping the quaternion-K\"ahler property
intact, as required by supersymmetry \cite{BW}. At the perturbative
level, a non-trivial one-loop correction modifies the low-energy tensor
multiplet Lagrangian \eqref{TM-action}, as was shown in \cite{AMTV}.
After dualization, this corrects the universal hypermultiplet metric
\eqref{class-UHM}, while still preserving the isometries
\eqref{Heis-alg}. More recently, this one-loop correction was written
and analyzed in the language of projective superspace in \cite{ARV},
using the tools developed in \cite{dWRV}.

At the non-perturbative level, there can be membrane and fivebrane
instantons. The latter were analyzed in \cite{DTV}. Membrane instantons,
which we are focussing on in this paper, arise from wrapping Euclidean
D2-branes around three-cycles in the CY$_3$ \cite{BBS}. For rigid
Calabi-Yau's, there are two kind of membrane instantons, as there are
two (supersymmetric) three-cycles that the membrane can wrap around.
Correspondingly, there will be two membrane instanton charges. These
instantons also have an effective supergravity description, as was shown
in \cite{TV1,DDVTV}. The two instanton charges correspond to the shift
symmetries in $\chi$ and $\varphi$, as written down in \eqref{Heis-alg}.
We denote these charges by $Q_\chi$ and $Q_\varphi$ respectively. They
can also be understood as being the charges of the corresponding dual
3-form field strengths that appear after dualizing one of the scalars
$\chi$ or $\varphi$ to a 2-form. Upon doing so, the tensor multiplet
becomes a double-tensor multiplet, in which the instanton solution can
be derived from a Bogomol'nyi equation \cite{TV1,DDVTV}. Following this
procedure, it becomes clear that only one charge can be switched on
simultaneously, either $Q_\chi$ or $Q_\varphi$, depending on which
scalar was dualized to a tensor. One cannot dualize both scalars to
tensors, as the two shift symmetries on $\chi$ and $\varphi$ do not
commute. In section \ref{CY}, we will rederive this property from a
string theory perspective.

The instanton action is inversely proportional to the string coupling,
which, in our conventions, is defined as
 \begin{equation} \label{gstring}
  g_s = \e^{-\phi_\infty/2}\ .
 \end{equation}
The membrane instanton action, say for the $\varphi$-instanton, then is
\cite{TV1,DDVTV}
 \begin{equation} \label{inst-action}
  S_\mathrm{inst} = 2\, \frac{|Q_\varphi|}{g_s} + \I \varphi\,
  Q_\varphi\ .
 \end{equation}
The imaginary term comes from a surface term that arises upon dualizing
the tensor to a scalar. It involves the zero mode of the dual scalar
$\varphi$, which can be identified with the value of the field at
infinity. Its presence breaks the shift symmetry in $\varphi$ to a
discrete subgroup. A similar formula also holds for the
$\chi$-instanton, by simply replacing $\varphi$ by $\chi$.
Notice also the factor $2$ in front of the real part of the instanton
action \eqref{inst-action}. This will become important later.

To compare, the NS-fivebrane instanton action is inversely proportional
to the square of the string coupling and, in the same normalization as
above, has no factor of 2 in front \cite{DTV}. It has a theta-angle-like
term proportional to the zero mode of $\sigma$. As long as we don't
switch on fivebrane instantons, the continuous shift symmetry in
$\sigma$ will remain an exact symmetry. In other words, in the absence
of fivebrane instantons, the quantum corrected universal hypermultiplet
moduli space will be a quaternionic manifold with a (non-compact) U(1)
isometry. Such manifolds have been classified by mathematicians in terms
of a single function, as we describe in the next section.

\section{Toda equation and universal hypermultiplet}
\label{PT}

As explained in the previous section, the effect of membrane instantons
is to modify the hypermultiplet moduli space non-perturbatively, in a
way consistent with the constraints from quaternion-K\"ahler (QK)
geometry. In the absence of fivebranes the quaternionic manifold has an
isometry that acts as a shift in the NS scalar $\sigma$. In this
section, we discuss the geometry of QK manifolds with a U(1) isometry,
and explain how the universal hypermultiplet fits into this framework.

\subsection{The Przanowski-Tod metric} 

In \cite{P} Przanowski derived the general form of four-dimensional
quaternion-K\"ahler metrics with (at least) one Killing vector. It was
later rederived by Tod \cite{T}. The Przanowski-Tod (PT) metric in local
coordinates $(r,u,v,t)$ reads
 \begin{equation} \label{1.1}
  \d s^2 = \frac{1}{r^2} \Big[ f \d r^2 + f \e^h (\d u^2 + \d v^2)
  + f^{-1} (\d t + \Theta)^2 \Big]\ .
 \end{equation}
The isometry acts as a shift in the coordinate $t$. The metric is
determined in terms of one scalar function $h(r,u,v)$, which is subject
to the three-dimensional Toda equation
 \begin{equation} \label{Toda}
  (\p_u^2 + \p_v^2) h + \p_r^2\, \e^h = 0\ .
 \end{equation}
The function $f(r,u,v)$ is not independent, but related to $h$ through
 \begin{equation} \label{f-h}
  f = - \frac{3}{2\Lambda} \big( 2 - r \p_r h \big)\ ,
 \end{equation}
while the 1-form $\Theta(r,u,v)=\Theta_r\d r+\Theta_u\d u+\Theta_v\d v$
is a solution to the equation
 \begin{equation} \label{dT}
  \d \Theta = (\p_u f\, \d v - \p_v f\, \d u) \wedge \d r
  + \p_r (f \e^h)\, \d u \wedge \d v\ .
 \end{equation}
Manifolds with such a metric are Einstein with anti-selfdual Weyl
tensor, and $\Lambda$ in \eqref{f-h} is the target space cosmological 
constant, $R_{AB}=\Lambda G_{AB}$.

As long as at least one isometry remains unbroken, the universal
hypermultiplet moduli space metric \eqref{class-UHM} is of this form.
Its Ricci tensor is found to be $R_{AB}=(-3/2)G_{AB}$, thus $\Lambda=
-3/2$ in our conventions.\footnote{More on our conventions on
quaternionic geometry can be found in appendix A\@.}

It is quite remarkable that the non-perturbative structure of the
universal hypermultiplet is fully encoded by the solutions of the Toda
equation. This equation has been studied by mathematicians in the
context of three-dimensional Einstein-Weyl spaces and hyperk\"ahler
manifolds \cite{BF,L,W} (see also appendix B). More recently, a large
class of solutions of the Toda equation was constructed by \cite{CT},
see also \cite{BS}. Unfortunately these do not seem to satisfy the
boundary conditions required by our set-up, so in the next section we
will construct new solutions that describe membrane instanton effects.

Integrable structures, including the Toda hierarchy, have also been
discovered in topological string theory \cite{ADKMV}. Related to this,
the Toda equation also appears in the non-perturbative description of
the non-critical $c=1$ string theory \cite{AK}. It would be interesting
to better understand the connection, if any, to our work. Finally, we
mention that the Toda equation also plays an important role in
classifying BPS vacua in M-theory \cite{LLM}.

\subsection{Symmetries, moment maps, and 4-fermion couplings} 

Clearly, the PT metric has a Killing vector $\p_t$ corresponding to a
shift symmetry in $t$. In coordinates $(r,u,v,t)$, this Killing vector
is given by
 \begin{equation} \label{k_t}
  k^A = (\, 0\, ,\, 0\, ,\, 0\, ,\, e_0\, )^\mathrm{T}\ ,\qquad e_0 \in
  \fieldR\ .
 \end{equation}
The moment maps of the shift symmetry can be computed from \eqref{2.9},
which we do in appendix \ref{appA.2}. The result is \emph{independent}
of the functions $f$, $h$ and $\Theta$, and reads
 \begin{equation} \label{eq:moment1}
  P^1 = 0\ ,\qquad P^2 = 0\ ,\qquad P^3 = \frac{e_0}{r}\ .
 \end{equation}

Furthermore, $4n$-dimensional quaternion-K\"ahler manifolds admit a
completely symmetric rank four tensor $\cW_{\alpha\beta\gamma\delta}$,
where $\alpha=1,...,2n$ labels the $\USp(2n)$ index that is part of the
holonomy group of QK manifolds. This tensor can be constructed out of
the Riemann curvature tensor; its definition and properties are
discussed in \cite{BCDGVV}, which we summarize in appendix \ref{appA.3}.
In $N=2$ supergravity effective actions, the $\cW$-tensor is contracted
with four hyperinos; it will play an important role in section \ref{CY}.
In our case, the QK manifold is four-dimensional and hence $n=1$.

For the PT metric we carry out its construction in appendix \ref{appA.3}
and state here only the final result:
 \begin{align} \label{r.2}
  \cW_{1111} & = 4 r^2 f^{-3} \e^{-h} \big[ f (\p^2_{\zb} f - \p_{\zb}
    h\, \p_{\zb} f) - 3 (\p_{\zb} f)^2 \big] \notag \\[2pt]
  \cW_{2111} & = r^2 f^{-3} \e^{-h/2} \big[ 2 f\, \p_r \p_{\zb} f - 3
    (f\, \p_r h + 2 \p_r f)\, \p_{\zb} f + f^2 \p_r \p_{\zb} h \big]
    \notag \\[2pt]
  \cW_{2211} & = - r^2 f^{-3} \big[ f \big( r \p^3_r h - (\p_r h)^2
    \big) - 4 \e^{-h}\, \p_z f\, \p_{\zb} f + 2 (\p_r f)^2 \big]
    \notag \\[2pt]
  \cW_{2221} & = - r^2 f^{-3} \e^{-h/2} \big[ 2 f\, \p_r \p_z f - 3
    (f\, \p_r h + 2 \p_r f)\, \p_z f + f^2 \p_r \p_z h \big] \notag
    \\[2pt]
  \cW_{2222} & = 4 r^2 f^{-3} \e^{-h} \big[ f (\p^2_z f - \p_z h\, \p_z
    f) - 3 (\p_z f)^2 \big]\ .
 \end{align}
Here we have introduced the complex variable $z=u+\I v$ in order to
write the components of $\cW_{\alpha\beta\gamma\delta}$ in a compact
way. We will use this tensor in a comparison of the properties of our
instanton corrected universal hypermultiplet metric with the results
for four-fermi correlation functions computed in string theory
\cite{BBS}.

\subsection{The universal hypermultiplet in the PT framework} 

To rewrite the metric \eqref{class-UHM} in the PT form, we have to
identify the moduli of the universal hypermultiplet with the PT
coordinates. This must be done consistently with the isometries, in
particular with the shift symmetry in the coordinate $t$. From the
Heisenberg algebra of isometries \eqref{Heis-alg} it is apparent that
one can choose to identify $t$ with either $\sigma$ or $\varphi$. The
shift symmetries are generated by the parameters $\alpha$ and $\beta$,
respectively. This leads to two `dual' representations of the PT metric
that describe the same moduli space. We can call these bases the
membrane and the fivebrane basis, respectively.

In the membrane basis, which is the relevant basis for our purposes, we
identify the coordinate $t$ with $\sigma$, such that the $\alpha$-shift
symmetry is manifest. This is because of the absence of fivebrane
instantons, which would break the continuous $\alpha$-shift symmetry to
a discrete subgroup \cite{DTV}. So, the coordinates can be chosen as
 \begin{equation}\label{3.8} 
  t=\sigma\ ,\qquad r = \e^\phi\ ,\qquad u = \chi\ ,\qquad v = \varphi\ .
 \end{equation}
In this basis, the classical moduli space metric of the universal
hypermultiplet corresponds to the solution $\e^h=r$ of the Toda equation
\eqref{Toda}. It follows that $f=1$ and $\Theta=u\,\d v$, the latter
being defined only modulo an exact form.

As mentioned above, besides the instanton contributions that we want to
determine in this paper, there are also perturbative quantum corrections
to the moduli space metric \cite{AMTV}. These can easily be incorporated
in our approach: Observe that with $h(r,u,v)$ also $h(r+c,u,v)$ is a
solution to the Toda equation for constant $c\in\fieldR$. Applied to the
classical solution $\e^h=r$, we obtain
 \begin{equation} \label{pertm}
  \e^h = r + c\ ,\qquad f = \frac{r + 2c}{r + c}\ ,\qquad \Theta =
  u\, \d v \ ,
 \end{equation}
which turns out to describe the 1-loop (in the string frame) corrected
metric of \cite{AMTV} if we identify
 \begin{equation}\label{defc}
  c = - \frac{4\, \zeta(2)\, \chi(X)}{(2\pi)^3} = - \frac{1}{6\pi}\,
  (h_{1,1} - h_{1,2})\ .
 \end{equation}
Here $h_{1,1}$ and $h_{1,2}$ are the Hodge numbers of the CY threefold
$X$ on which the type IIA string has been compactified; for rigid CY's,
where $h_{1,2}=0$, we have the important bound $c<0$. The function $h$
in \eqref{pertm} is simply the general $(u,v)$-independent solution to
the Toda equation (modulo a constant rescaling of $r$); in this sense
the perturbative corrections appear naturally. The PT coordinate $r$
is related to $\rho$ in \cite{AMTV} through $r=\rho^2-c=\e^\phi$; the
relation between the fields and PT coordinates receives no
(perturbative) quantum corrections.

Note that if we consider $c<0$, the function $f$ in \eqref{pertm}
becomes negative for $r<2|c|$, which results in a negative-definite
metric \eqref{1.1}. We thus have to restrict $r$ to the open interval
$2|c|<r<\infty$.

\section{Instanton corrections} \label{sect4}

In this section, we construct solutions to the Toda equation
\eqref{Toda} that include an (infinite) series of exponential
corrections describing the membrane instantons. As we have learned from
the supergravity description, the real part of the instanton action is
inversely proportional to the dilaton, which becomes the square root
of the radial variable $r$. The precise form of the supergravity
instanton action is given in \eqref{inst-action}. This motivates us to
make a general ansatz of the form
 \begin{equation} \label{Ansatz}
  \e^h = r + \sum_{n\geq 1}\, \sum_m\, f_{n,m}(u,v)\, r^{-m/2+\alpha}\,
  \e^{-2n\sqrt{r}}\ .
 \end{equation}
As explained in the previous section, one can shift the value of $r$
with a constant to construct a new solution. This will then include the
perturbative one-loop correction of \cite{AMTV}. The power series in $r$
in front of the exponent describes the perturbative corrections around
the instantons. Using \eqref{gstring} we have that $r^{-m/2}=g_s^m$, and
the sum over $m$ is over the integers $\fieldZ$. At each instanton level
$n$, there is a lowest value $m_n$ that defines the leading term in the
expansion,
 \begin{equation}
  f_{n,m}(u,v) = 0 \quad \text{for} \quad m < m_n\ .
 \end{equation}
We have also introduced a parameter $\alpha$ which, without loss of
generality, lies in the interval $[0,1/2[$. This leaves open the
possibility that the leading term is not an integer power of $g_s$, as
e.g.\ in \cite{OV}. We will show later on that the Toda equation
enforces $\alpha=0$. With the $r$-dependence made explicit, solving the
Toda equation amounts to solving the differential equations for the
functions $f_{n,m}(u,v)$. These are of the type of inhomogeneous
Laplace equations, and we can solve them iteratively, order by order
in $n$ and $m$, to any order needed.

To get some additional insight, we focus for a moment on the asymptotic
(large $r$) behavior of the solution. We can then further specify the
ansatz as
 \begin{equation} \label{as-sol}
  \e^h = r + A \cos(k_uu+k_vv)\, r^{\beta} \e^{-2k\sqrt{r}}\ ,
 \end{equation}
with $A$ a normalization constant. One can now check that, to leading
order, the Toda equation is satisfied for any value of $\beta$, provided
that
 \begin{equation} \label{k-rel}
  k^2 = k_u^2 + k_v^2\ .
 \end{equation}
This asymptotic behavior indeed reproduces leading order charge $k$
instanton effects, including a one-loop correction in front of the
exponent. The cosine in the ansatz \eqref{as-sol} could also be
replaced by a sine, or a linear combination. Rewriting them in terms
of exponentials, one produces theta-angle like terms for both
instantons and anti-instantons, depending on the signs of $(k_u,k_v)$.
The relation \eqref{k-rel} is completely consistent with the
supergravity description of the instanton action \eqref{inst-action},
which describes the special case of either $k_u=0$ or $k_v=0$.

We now give a more complete analysis for solving the Toda equation,
based on the general ansatz \eqref{Ansatz}. This will enable us to
determine the subleading corrections to the solution \eqref{as-sol}.
More technical details are given in appendix C. For instance, in
appendix \ref{AppC1} we show that $m_n\geq -2$ for all $n$, and in
appendix \ref{AppC2} it is proven that $\alpha=0$.

We first bring the Toda equation into the equivalent form
 \begin{equation} \label{Toda2}
  \e^h \big( \p_u^2 + \p_v^2 + \e^h \p_r^2 \big) \e^h - (\p_u \e^h)^2
  - (\p_v \e^h)^2 = 0\ ,
 \end{equation}
such that it depends on $h$ only through $\e^h$. We then decompose
this equation into $N$-instanton sectors, each containing a sum over
all loop corrections,
 \begin{align}\label{4.5}
  0 = \sum_{n,m} &\ r^{-m/2}\, \e^{-2n\sqrt{r}}\, \Big\{ (\Delta +
    n^2)\, f_{n,m+2} + n\, a_{m+2}\, f_{n,m+1} + b_{m+2}\, f_{n,m}
    \notag \\[-2pt]
  & + \sum_{n',m'} \e^{-2n'\!\sqrt{r}}\, \big[ 2n\, a_{m'+1}\, f_{n',
    m-m'-1} + 2 b_{m'+2}\, f_{n',m-m'-2} \notag \\[-6pt]
  & \mspace{112mu} + f_{n',m-m'}\, (\Delta + 2n^2) - \nabla f_{n',m-m'}
    \cdot \nabla \big] f_{n,m'} \notag \\[4pt]
  & + \sum_{n',m'} \sum_{n'',m''} \e^{-2(n'+n'')\sqrt{r}}\, f_{n,m'}
    f_{n',m''}\, \big[ n^2 f_{n'',m-m'-m''-2} \notag \\[-8pt]
  & \mspace{94mu} +  n\, a_{m'+1}\, f_{n'',m-m'-m''-3} + b_{m'+2}\,
    f_{n'',m-m'-m''-4} \big] \Big\}\ ,
 \end{align}
where $\nabla=(\p_u,\p_v)$, $\Delta=\nabla^2$, and
 \begin{equation}\label{4.6}
  a_m = \half\, (2m-1)\ ,\qquad b_m = \quart\, m(m-2)\ .
 \end{equation}
In the $(N=1)$-instanton sector only the single-sum terms contribute,
while the double- and triple-sums have to be taken into account
beginning with the $(N=2)$- and $(N=3)$-instanton sectors, respectively.

\subsection{The one-instanton sector} 

We start with $N=1$. In this sector the Toda equation requires at the
$m$th loop order
 \begin{equation} \label{f1m}
  (\Delta + 1)\, f_{1,m} + a_m\, f_{1,m-1} + b_m\, f_{1,m-2} = 0\ ,
 \end{equation}
It is convenient to first consider a one-dimensional truncation, where
$f_{1,m}=f_{1,m}(x)$ with $x\in\{u,v\}$. In appendix \ref{AppC3} we
prove that the general one-dimensional solution of \eqref{f1m} is given
by
 \begin{equation} \label{solf1m}
  f_{1,m}(x) = \re \sum_{s\geq 0}\, \frac{1}{s!\,(-2)^s}\, k_{1,m}(s)\,
  G_s(x)
 \end{equation}
with recursively defined coefficients
 \begin{equation}\label{recurse-rel}
  k_{1,m}(s+1) = a_m k_{1,m-1}(s) + b_m k_{1,m-2}(s)\ .
 \end{equation}
$G_s(x)$ are complex functions related to the spherical Bessel functions
of the third kind; their precise definition can be found in appendix
\ref{AppC3}. $k_{1,m}(0)=A_{1,m}$ are complex integration constants
originating from the homogeneous part of \eqref{f1m}. By definition of
$m_1$ we have that $A_{1,m}=0$ for $m<m_1$, and from this it follows
that $k_{1,m}(s>m-m_1)=0$ by using \eqref{recurse-rel}. The highest
$x$-monomial contained in $f_{1,m}(x)$ is then of order $m-m_1$.
Explicitly, the first two solutions read
 \begin{align}
  f_{1,m_1}(x) & = \re \big\{ A_{1,m_1} \e^{\I x} \big\}\, , \notag
    \\[2pt]
  f_{1,m_1+1}(x) & = \re \big\{ A_{1,m_1+1} \e^{\I x} + \half a_{m_1
    +1} A_{1,m_1}\, \I x\, \e^{\I x} \big\}\ . \label{solf1m0}
 \end{align}

We now extend the $N=1$ result to the general $u,v$ dependent solution.
This can be done by Fourier transforming in the $u,v$ plane or, as we do 
below, by separation of variables. In both cases one finds a basis of 
solutions; the most general solution is then obtained by superposition.
Using separation of variables, we find a basis and parameterize it by
a continuous parameter $\lambda$.

Introducing $\omega=\sqrt{1-\lambda^2}$ with $\lambda^2$ being real,
the general solution can then be written as 
 \begin{gather}
  f_{1,m}(u,v) = \int\! d\lambda\, \re \sum_{s\geq 0}\, \frac{1}{s!\,
    (-2\, \omega^2)^s}\, k_{1,m}(s, u; \lambda)\, G_s(\omega v)\ ,
    \notag \\[2pt]
  k_{1,m}(s+1, u; \lambda) = a_m k_{1,m-1}(s, u; \lambda) + b_m k_{1,
    m-2}(s, u; \lambda)\ , \label{solf2m}
 \end{gather} 
where
 \begin{equation}
  k_{1,m}(0, u ; \lambda) = B_{1,m}(\lambda)\, A_{1,m}(\lambda)\,
  \e^{\I\lambda u}\ .
 \end{equation}
Here $A_{1,m}(\lambda)$, $B_{1,m}(\lambda)$ are arbitrary complex
integration functions which determine the ``frequency spectrum'' of the
solution. The $u$-independent solution of the previous paragraph can
then be obtained by setting
 \begin{equation}
  A_{1,m}(\lambda) = A_{1,m}\ ,\qquad B_{1,m}(\lambda) = \delta
  (\lambda)\ ,
 \end{equation}
with $A_{1,m}$ being the corresponding integration constants.

One can now combine the general $u$-independent solution with the
general $v$-inde\-pen\-dent solution. This can be done by taking the
coefficient functions $A_{1,m}(\lambda)$ and $B_{1,m}(\lambda)$ to be
peaked around $\lambda=0$ and $\lambda=1$. This is not
the most general solution, but it is the preferred one that describes
our physical problem. For general values $0<\lambda<1$, one generates
products of exponents in $u$ and in $v$ that describe theta-angle-like
terms in the supergravity instanton action where both $\varphi$ and
$\chi$ and their charges $Q_\varphi$ and $Q_\chi$ are turned on. As we
have argued at the end of section \ref{sugra-des}, this cannot be the
case. Moreover, as we will see in section \ref{CY}, string theory also
predicts such terms to be absent. We therefore only take contributions
from $\lambda=0,1$. This implies that $\e^h$, including the perturbative
corrections and the instanton corrections arising in the one-instanton
sector, can be completely expressed in terms of the one-dimensional
solutions \eqref{solf1m}. Here making the substitution $x\rightarrow
u,v$ describes the one-instanton contribution to $\e^h$ arising from a
$u,v$-instanton, respectively. Also taking into account the perturbative
corrections to the solution by shifting $r \rightarrow r+c$ we find
 \begin{equation} \label{9.1}
  \exp [h(r,u,v)] = \exp [h_\text{pert}(r)] + \exp [h_\text{1-inst}(r,u)
  ] + \exp [h_\text{1-inst}(r,v)] + \ldots\ ,
 \end{equation}
where
\begin{equation}\label{9.2}
  \exp [h_\text{pert}(r)] = r + c \ ,\qquad
  \exp [h_\text{1-inst}(r,u)]  = \e^{-2 \sqrt{r+c}}\, \sum_{m\ge m_1}
	f_{1,m}(u)\, (r+c)^{-m/2} \ ,
\end{equation}
and similarly for $h_\text{1-inst}(r,v)$. The coefficients $f_{1,m}(x)$
are the one parameter solution \eqref{solf1m} and the ellipses denote
the contributions from higher order instanton corrections.
 
For later reference, we also give the leading order expression for
$\e^h$ in the regime $r\gg 1$ (small string coupling). To leading order
in the semi-classical approximation, the instanton solution \eqref{9.1}
reads
 \begin{equation} \label{eh-exp}
  \e^h = r + c + \frac{1}{2}\, r^{-m_1/2}\, \big( A_{1, m_1}\, \e^{\I v}
  + A_{1, m_1}^*\, \e^{-\I v} + B_{1, m_1}\, \e^{-\I u} + B_{1, m_1}^*\,
  \e^{\I u} \big)\, \e^{-2\sqrt{r}} + \ldots\ .
 \end{equation}
Notice that we need to include both instantons and anti-instantons to
obtain a real solution.

To find the leading-order instanton corrected hypermultiplet metric, we
first compute the leading corrections to $f$ defined in $\eqref{f-h}$:
 \begin{equation} \label{finst}
  f = \frac{r + 2c}{r+c} + \frac{1}{2}\, r^{-(m_1 + 1)/2}\, \big( A_{1,
  m_1}\, \e^{\I v} + A_{1,m_1}^*\, \e^{-\I v} + B_{1,m_1} \, \e^{-\I u}
  + B_{1,m_1}^*\, \e^{\I u} \big)\, \e^{-2\sqrt{r}} + \ldots\ . 
 \end{equation}
Substituting this result into \eqref{dT}, one derives the leading
corrections to the $\Theta$ 1-form. Setting
 \begin{equation}
  \Theta = u\, \d v\ + \Theta_\text{inst}\ ,
 \end{equation}
these are given by
\begin{equation}\label{Tinst}
 \Theta_\text{inst} = r^{-m_1/2}\, \e^{-2 \sqrt{r}}\, \im \{ A_1\,
  \e^{\I v}\, \d u + B_1\, \e^{-\I u}\, \d v \}\ + \ldots\ .
\end{equation}
The leading order corrections to the hypermultiplet scalar metric are
then obtained by plugging these expressions into the PT metric
\eqref{1.1}.

\subsection{Higher instanton sectors} 

We now briefly discuss the $N=2$ sector. The Toda equation requires at
this level
 \begin{align}\label{N=2sector}
  0 & = (\Delta + 4)\, f_{2,m} + 2a_m\, f_{2,m-1} + b_m\, f_{2,m-2}
    \notag \\[2pt]
  & \tab + \sum_{m'} \big[ f_{1,m-m'-2} + a_{m'+1}\, f_{1,m-m'-3}
    + b_{m'+2}\, f_{1,m-m'-4} \notag \\[-6pt]
  & \mspace{76mu} - \nabla f_{1,m-m'-2} \cdot \nabla \big] f_{1,m'}\ ,
 \end{align}
where we have used \eqref{f1m} for $\Delta f_{1,m'}$ in the double sum.
We have not derived the general solution to these equations in closed
form; the one-dimensional truncation, however, is straightforward to
solve order by order in $m$. At lowest order\footnote{In appendix
\ref{AppC2} we show that for $n\geq n'$ it is $-2\leq m_n\leq m_{n'}$.}
$m_2$ we have
 \begin{equation} \label{2m0eq}
  (\Delta + 4)\, f_{2,m_2} + \de_{m_2,-2}\, \big[ (f_{1,m_2})^2 -
  (\nabla f_{1,m_2})^2 \big] = 0\ .
 \end{equation}
Note that the inhomogeneous term is present only for the lowest possible
value $m_2=-2$. The one-dimensional truncation yields the equation
 \begin{equation}
  (\p_x^2 + 4)\, f_{2,m_2}(x) + \de_{m_2,-2}\, \re \big\{ A_{1,m_2}^2
  \e^{2\I x} \big\} = 0\ ,
 \end{equation}
where we have inserted the solution \eqref{solf1m0} for $f_{1,m_1}(x)$.
The general solution then reads
 \begin{align}
  f_{2,m_2}(x) & = \re \big\{ A_{2,m_2} \e^{2\I x} + \tfrac{1}{4}
    \de_{m_2,-2}\, A_{1,m_2}^2\, \I x\, \e^{2\I x} \big\} \notag
    \\[2pt]
  & = \re \big\{ A_{2,m_2} G_0(2x) - \tfrac{1}{8} \de_{m_2,-2}\, A_{1,
    m_2}^2 G_1(2x) \big\}\ ,
 \end{align}
$A_{2,m_2}$ being a further complex integration constant. 

The solution for $m>m_2$ can now be constructed by solving the
appropriate equation arising from \eqref{N=2sector}. Based on
\eqref{solf2m} we can also construct the general $(u,v)$-dependent
solution for $f_{2,m_2}(u,v)$. The idea is to decompose the products of
$\cos(\lambda_1 u)\,\cos(\lambda_2 u)$, etc., appearing in the
inhomogeneous part of \eqref{2m0eq} into a sum of $\cos$ and $\sin$
terms using product formulae for two trigonometric functions. We can
then construct the full inhomogeneous solution by superposing the
inhomogeneous solutions for every term in the sum. We refrain from
giving the result, however, since it is complicated and not particularly
illuminating.

We conclude this subsection by giving an argument that the iterative
solution devised above indeed gives rise to a consistent solution of the
Toda equation. The general equations which determine a new $f_{n,m}(u,
v)$ are two-dimensional Laplace equations to the eigenvalue $n^2$
coupled to an inhomogeneous term, which is completely determined by the
$f_{n,m}(u,v)$'s obtained in the previous steps of the iteration
procedure. These equations are readily solved, e.g., by applying a
Fourier transformation. It then turns out that the iteration procedure
is organized in such a way that any level in the perturbative expansion
\eqref{4.5} determines one ``new'' $f_{n,m}(u,v)$, i.e., there are no
further constraints on the $f_{n,m}(u,v)$ determined in the previous
steps. This establishes that our perturbative approach indeed extends to
a consistent solution of the Toda equation \eqref{Toda}.

\subsection{The fate of the Heisenberg algebra} 
\label{sect4.3}

Based on the Toda solution \eqref{9.1} we now discuss the breaking of
the Heisenberg algebra \eqref{Heis-alg} in the presence of membrane
instantons. We start with the shift symmetry in the axion $\sigma
\rightarrow\sigma-\alpha$. By identifying $t=\sigma$, this shift 
corresponds to the isometry of the Tod metric, so that it cannot
be broken by the instanton corrections.

Analyzing the $\beta$ and $\gamma$-shifts is more involved. Under the
identification \eqref{3.8}, the $\beta$-shift then acts as $v\rightarrow
v+\beta$. Taking the leading order one-instanton solution
\eqref{eh-exp}-\eqref{Tinst}, we find that $\e^h$ as well as the
resulting functions $f$ and $\Theta$ appearing in the metric depend on
$v$ through $\e^{\pm\I v}$ or $\d v$ only. These theta-angle-like terms
break the $\beta$-shift to the discrete symmetry group
$\fieldZ$.\footnote{This agrees with earlier observations made in
\cite{BB}.} Going beyond the leading instanton corrections by taking
into account higher loop corrections around the single instanton will,
however, generically break the $\beta$-shift completely, due to the
appearance of polynomials in $v$ multiplying the factors $\e^{\pm\I v}$.
We point out, however, that by setting the integration constants
multiplying the terms odd in $v$ to zero, there is still an unbroken
$\fieldZ_2$ symmetry defined by $v\rightarrow-v$, $t\rightarrow-t$, 
interchanging $v$-instantons and anti-instantons.

To deduce the fate of the $\gamma$-shift, $u\rightarrow u+\gamma$,
$t\rightarrow t-\gamma v$, we first observe that $t\rightarrow t-\gamma
v$ implies that the combination $\d t+u{\rm dv}$ is invariant.
Applying the same logic as for the $\beta$-shift above, we then find
that the one-loop corrections of a single $u$-instanton break the
$\gamma$-shift to the discrete symmetry $\fieldZ$, which will be
generically broken by higher order terms appearing in the loop
expansion. Similar to the $\beta$-shift, however, we can arrange the
constants of integration appearing in the solution in such a way that
there is also a $\fieldZ_2$ symmetry. We expect that these two
$\fieldZ_2$ symmetries could play a prominent role when determining
(some of) the coefficients appearing in the solution \eqref{9.1} from
string theory.

\section{Comparison to string theory} \label{CY}

As mentioned above, instanton corrections to the moduli space metric
also induce corrections to the 4-fermion couplings in the supergravity
effective action, since they couple to the curvature of the moduli space
metric. It is therefore desirable to have a microscopic string theory
derivation that reproduces these instanton corrections. Using the work
of Becker, Becker and Strominger (BBS) \cite{BBS}, this is possible, and
we show in this section that there is a perfect agreement with string
theory.

The reason why 4-fermi terms are the relevant objects to look at is that
our membrane instantons break half of the supersymmetries. The resulting
four fermionic zero modes then lead to non-vanishing 4-fermion
correlation functions. This was already observed by BBS in a string
theoretic setting. Here we compare our supergravity result for the
instanton corrected 4-hyperino couplings with those derived in
\cite{BBS}. Their analysis was actually set up by starting with CY$_3$
compactifications of M-theory, and then reducing to type IIA in ten
dimensions. As we also explain below, the only modifications are in the
appearance of the string coupling constant. This is also consistent
with the supergravity analysis, since the hypermultiplet couplings to
supergravity are (almost) identical in four and five space-time
dimensions.

\subsection{The string calculation} 

The relevant curvature tensor that is contracted with the 4-fermi terms
is the totally symmetric $\cW_{\alpha\beta\gamma\delta}$ tensor
introduced in section 3.2 (see also appendix \ref{appA.3}). In the BBS
paper, this tensor was denoted by $\cR_{IJKL}$. For compactifications on
rigid CY$_3$ yielding one (the universal) hypermultiplet, we expect that
they agree up to normalization (which, to our knowledge, has not been
computed in a string theory setting) and an $\USp(2)\simeq\SU(2)$
rotation of the fermion frame.

Instanton configurations are obtained by wrapping Euclidean membranes
over a supersymmetric three-cycle $\cC_3$. The effect of such an
instanton is to yield a non-vanishing 4-fermi correlator that gives a
contribution to the curvature tensor. In the M-theory set-up, this was
found to be (see eq.\ (2.49) in \cite{BBS}),
 \begin{equation} \label{S1}
  \Delta_{\cC_3} \cR_{IJKL} = N'\, \e^{-S_\inst}\, \int_{\cC_3}\! d_I\,
  \int_{\cC_3}\! d_J\, \int_{\cC_3}\! d_K\, \int_{\cC_3}\! d_L\ .
 \end{equation}
Here, $N'$ is an unspecified normalization factor which, in principle,
could depend on the string coupling $g_s$. Furthermore, 
 \begin{equation} \label{S2}
  S_\inst = \e^{-\cK}\, \Big| \int_{\cC_3}\! \Omega\, \Big| + \I
  \int_{\cC_3}\! C_3
 \end{equation}
is the bosonic part of the instanton action, $\cC_3$ denotes the
supersymmetric cycle that is wrapped by the membrane, $C_3$ is the
3-form potential in $D=11$ supergravity, and the $d_I$ form a real
symplectic basis of $H^{3}(X,\fieldZ)$. Finally, we have $\cK=1/2\,
(\cK_V-\cK_H)$ with
 \begin{equation}
  \cK_V = - \log \left( \frac{4}{3} \int_X\! \hat{J} \wedge \hat{J}
  \wedge \hat{J} \right)\ ,\qquad \cK_H = - \log \left( \I \int_X\!
  \Omega \wedge \bar{\Omega} \right)\ ,
 \end{equation}
where $\hat{J}$ is the K\"ahler form and $\Omega$ the holomorphic 3-form
on the CY threefold $X$.

All these quantities can be expressed in terms of our variables for the
universal hypermultiplet. For this we need the relation between the Tod
variables $(r,u,v,t)$ and the fields appearing in the IIA superstring
action of BBS \cite{BBS}. To establish these relations the references
\cite{LM,GLMW} are useful.

In order to compactify the IIA string on a CY$_3$ manifold $X$, we
introduce $2(h_{1,2}+1)$ harmonic 3-forms $(\alpha_a, \beta^a)$, which
form a real basis of $H^3(X,\fieldZ)$, with the usual normalization
 \begin{equation} \label{S3}
  \int_X \alpha_a \wedge \beta^b = - \int_{X} \beta^b \wedge \alpha_a =
  \delta^b_a\ ,\qquad \int_X \alpha_a \wedge \alpha_b = \int_X \beta^a
  \wedge \beta^b = 0\ . 
 \end{equation}
They correspond to the $d^I$ in \eqref{S1}. Furthermore, we introduce
the canonical dual basis of real 3-cycles $(\cA^a,\cB_a)$ of $H_3(X,
\fieldZ)$, satisfying~\footnote{In these relations, we have chosen a 
normalization in which the volume of the CY$_3$ is set to one.}
 \begin{equation}\label{intnormal}
  \int_{\cA^a} \alpha_b = - \int_{\cB_b} \beta^a = \delta_b^a\ ,\qquad
  \int_{\cA^a} \beta^b  = \int_{\cB_a} \alpha_b = 0\ .
 \end{equation}
For rigid CYs, the index $a$ takes only the value $0$ and may be
omitted. We can then use this basis to define the periods of the
holomorphic 3-form $\Omega$ of the CY$_3$ as
 \begin{equation}
  z^a = \int_{\cA^a}\! \Omega\ ,\qquad \cG_a = \int_{\cB_a}\! \Omega\ ,
 \end{equation}
in terms of which
 \begin{equation}
  \Omega = z^a\, \alpha_a - \cG_a(z)\, \beta^a\ .
 \end{equation}
Here $z^a$ are the complex structure moduli, and $\cG_a(z)$ are
derivatives of the prepotential of the special geometry which is
parameterized by the $z^a$. The K\"ahler potential on the space of
complex structure deformations is then given by
 \begin{equation}
  \cK_H = - \log \big(\! -2 \im (\zb^a \cG_a) \big)\ .
 \end{equation}
In the case of a rigid CY$_3$ we only have $z^0$ and $\cG_0$. We can
then choose the normalization of $\Omega$ (which is defined up to a
complex rescaling only) such that
 \begin{equation}
  z^0 = 1\ ,\qquad \cG_0 = - \I\ .
 \end{equation}
The phase of $\cG_0$ is determined in such a way that $\cK_H$ is real.

With these prerequisites it is now possible to determine the
supersymmetric cycles of the CY$_3$\@. In fact, we shall find that
these are given by the cycles $\cA^a$, $\cB_a$ themselves. In
\cite{BBS} it was shown that a supersymmetric cycle has to satisfy
the following two (equivalent) conditions:
 \begin{enumerate}
  \item The pull-back of the embedding space's K\"ahler form has to
    vanish.
  \item The cycle has to be calibrated with respect to the holomorphic
    3-form $\Omega$, i.e., the volume form of the cycle has to be
    proportional to the pull-back of $\Omega$ up to a complex phase
    factor. (This implies that a supersymmetric cycle has to be a
    special Lagrangian submanifold of $X$ \cite{JG}.)
 \end{enumerate}

In order to show that the cycles $\cA^a$, $\cB_a$ satisfy the first
condition, we can generalize the argument given in the example of
\cite{BBS}, section 2.2. There, an isometry $D$ of the metric was
employed which corresponds to complex conjugation. The K\"ahler metric
on $X$ is invariant under $D$, as are the real cycles $\cA^a$, $\cB_a$,
whereas the K\"ahler form associated with the K\"ahler metric reverses
its sign,
 \begin{equation}
  D:\ \hat{J} \mapsto - \hat{J}\ .
 \end{equation}
On the other hand, the pullback of $\hat{J}$ onto $\cA^a$ or $\cB_a$,
respectively, must be invariant, which is only possible
if $\hat{J}$ vanishes on $\cA^a$ or $\cB_a$.

The second condition is satisfied in rigid CY$_3$ compactifications
since $\alpha$ and $\beta$ correspond to the induced volume forms on
$\cA$ and $\cB$, respectively.

We can now use the dimensional reduction outlined in \cite{LM,
GLMW}\footnote{The various fields in \cite{BBS}, \cite{LM,GLMW} and our
paper, respectively, differ in their normalizations. The UHM variables
in \cite{GLMW} are related to ours through $\hat{\phi}=-\phi/2$,
$\hat{\xi}=\varphi/\sqrt{2}$, $\hat{\bar{\xi}}=\chi/\sqrt{2}$, $\hat{a}
=\sigma+\half\chi\varphi$. The RR fields in \cite{BBS} are obtained
from those in \cite{GLMW} by multiplication with $\sqrt{2}$.} to
evaluate the integrals appearing in \eqref{S1}. We find that $C_3$ in
\cite{BBS} is expanded as
 \begin{equation}
  C_3 = c_3 + v\, \alpha + u\, \beta + \dots\ ,
 \end{equation}
where $c_3$ is the space-time 3-form potential (which is non-dynamical
in four dimensions) and the ellipses denote the omitted vector multiplet
contributions.

To evaluate the K\"ahler potential appearing in \eqref{S2}, we note that
the volume of the CY$_3$ manifold measured with the 11-dimensional
supergravity metric is given by
 \begin{equation}
  \hat{V}_6 = \frac{1}{3!}\, \int_X\! \hat{J} \wedge \hat{J} \wedge
  \hat{J}\ .
 \end{equation}
Upon reducing to the IIA supergravity action in the string frame using
 \begin{equation}
  \d \hat{s}^2_{11} = \e^{-2 \phi / 3} \big( \d x_{11} + A_m \d
  x^m \big)^2 + \e^{\phi/3} \d s^2_{10}\ ,
 \end{equation}
$\cK$ acquires a non-trivial dependence on the dilaton. Since
 \begin{equation}
  \hat{J} = \I\, \hat{g} = \e^{\phi / 3}\, \I\, g = \e^{
  \phi / 3}\, J\ ,
 \end{equation}
where $J$ is the K\"ahler form in the string frame, we find the relation
$\hat{V}_6=\e^{\phi}\,V_6$. In our normalization \eqref{S3},
 \begin{equation}
  \I\! \int_X\! \Omega \wedge \bar{\Omega} = 2 V_6\ ,
 \end{equation}
we can evaluate the K\"ahler potential in \eqref{S2}:
 \begin{align}
  \exp(-\cK) & = \exp \Big(\! - \frac{1}{2}\, (\cK_V - \cK_H) \Big)
	\notag \\
  & = \exp \Big(\! - \frac{1}{2}\, \big( -\log (8 \e^{\phi}\, V_6) +
	\log (2\, V_6) \big) \Big) \notag \\[2pt]
  & = 2\, \e^{\phi/2}\ .
 \end{align}
It is now also straightforward to evaluate the remaining integral
 \begin{equation}
  \Big| \int_{\cC_3}\! \Omega\, \Big| = \sqrt{\smash[b]{m^2 + n^2}}\ ,
 \end{equation}
for $\cC_3=m\cA+n\cB$. Notice, however, that under the condition that
$\cA$ and $\cB$ are calibrated one can show that the linear combination
$\cC_3=m\cA+n\cB$ is a calibrated cycle if and only if either $m\in
\fieldZ$, $n=0$ or $m=0$, $n\in\fieldZ$. Therefore a membrane wrapping
$\cA$ and $\cB$ simultaneously is not a supersymmetric configuration
and does not contribute to the instanton corrected metric. This is also
reflected in the form of $S_\text{inst}$ given in \cite{BB}, where
instanton charges are linear: $n+m$, where $m$ and $n$ cannot be
non-zero simultaneously.

Putting everything together, using $r=\e^\phi$, we then obtain the
instanton weight of a configuration where $\cC_3=m\cA+n\cB$:
 \begin{equation}
  \e^{-S_\mathrm{inst}} = \e^{-2 \sqrt{\smash[b]{m^2 + n^2}} \sqrt{r}}\,
  \e^{-\I m v + \I n u}\ . 
 \end{equation}
In the framework of a rigid CY$_3$ compactification, the $d^I$, $I=1,2$, 
correspond to the
harmonic three-forms $\alpha$, $\beta$, while the wrapped cycle $\cC_3$
can be either the cycle $\cA$ or $\cB$ introduced above. Using the
relations \eqref{intnormal} it is then straightforward to verify that
the instanton corrections predicted by \eqref{S1} enter into the
components of $\Delta_{\cC_3}\cR_{IJKL}$ with $I=J=K=L=1,2$ only.

As we will show in the next subsection, at leading order in $m$, $n$
this agrees with the results obtained by substituting our instanton
corrected UHM into $W_{\alpha\beta\gamma\delta}$ and also fixes the
coefficient functions in our solution in terms of free parameters.

\subsection{Comparison with the instanton corrected PT metric} 

Evaluating $\cW_{\alpha\beta\gamma\delta}$ for the universal
hypermultiplet in the membrane base, we find that at the perturbative
level (classical plus loop corrections) the only non-vanishing component
is given by
 \begin{equation}
  \cW_{2211} = - \frac{r^3}{(r+2c)^3}\ .
 \end{equation}
We now substitute the instanton expansion \eqref{eh-exp} in the general
$\cW$-tensor given in \eqref{r.2}. After subtracting the classical
contribution, we obtain to lowest order in $g_s$
 \begin{align} \label{r.3}
  \Delta \cW_{1111} & = N\, \big[ A_1\, \e^{\I v} + A_1^*\, \e^{-\I v}
    - B_1\, \e^{-\I u} - B_1^*\, \e^{\I u} \big] \notag \\[2pt]
  \Delta \cW_{1112} & =  N\, \big[ A_1\, \e^{\I v} - A_1^*\, \e^{-\I v}
    + \I (B_1\, \e^{-\I u} - B_1^*\, \e^{\I u}) \big] \notag \\[2pt]
  \Delta \cW_{1122} & =  N\, \big[ A_1\, \e^{\I v} + A_1^*\, \e^{-\I v}
    + B_1\, \e^{-\I u} + B_1^*\, \e^{\I u} \big] \notag \\[2pt]
  \Delta \cW_{1222} & = N\, \big[ A_1\, \e^{\I v} - A_1^*\, \e^{-\I v}
    - \I (B_1\, \e^{-\I u} - B_1^*\, \e^{\I u}) \big] \notag \\[2pt]
  \Delta \cW_{2222} & = N\, \big[ A_1\, \e^{\I v} + A_1^*\, \e^{-\I v}
    - B_1\, \e^{-\I u} - B_1^*\, \e^{\I u} \big]\ .
 \end{align}
Here we have set $N=(r+c)^{(1-m_1)/2}\,\e^{-2\sqrt{r+c}}$. Comparing the
$r$-dependence of $N$ with the one appearing in the normalization factor
$N'$ then fixes the value of $m_1$. In particular, for $N'$ being an
$r$-independent normalization constant, we obtain $m_1=1$.

At first sight the tensorial structure of \eqref{r.3} seems to disagree
with \eqref{S1}, as the latter predicts instanton corrections to the
components $\Delta\cW_{1111}$ and $\Delta\cW_{2222}$ only. In order to
match these two results it is crucial to observe that \eqref{S1}
describes the corrections arising from a single membrane wrapping
\emph{one} supersymmetric cycle, while our expression already contains
the ``instanton sum'' over the $\cA$ and $\cB$ cycle. Furthermore, the
fermion frame used by BBS does not necessarily agree with ours. However,
these two frames can differ at most by a local SU(2)
rotation\footnote{The rotation group has to be compatible with the
reality condition imposed on the pair of symplectic Majorana spinors
coupling to $\cW$ and preserve fermion bilinears. These conditions then
lead to the fact that the most general transformation is given by SU(2).
This is discussed in \cite{CMMS1}.}. We parameterize this transformation
by
 \begin{equation}
  U = \lp \e^{\I\xi} \cos\eta & \e^{\I\rho} \sin\eta \\ -\e^{-\I\rho}
  \sin\eta & \e^{-\I\xi} \cos\eta \rp\ ,
 \end{equation}
where the parameters $\eta$, $\xi$, $\rho$ can in principle depend on
the scalars (this, however, will not be necessary). To follow BBS, we
then consider the contribution arising from the $\cB$-instanton only.
This requires setting $A_1=0$. Upon performing a global $\SU(2)$
rotation of the fermion frame with parameters $\eta=\pi/4$, $\xi=\rho+
\pi/2$, we obtain
 \begin{equation} \label{r.4}
  \Delta \tilde{\cW}_{1111} = - 4 N B_1^*\, \e^{\I u}\ ,\qquad
  \Delta \tilde{\cW}_{2222} = - 4 N B_1\, \e^{-\I u}\ ,
 \end{equation}
with the other components vanishing identically. (Note that the
remaining free parameter in the transformation $\rho$ only induces a
phase on the components of $\Delta\tilde{\cW}_{\alpha\beta\gamma
\delta}$, which we set to zero for convenience.) Observe that this now
matches the prediction of BBS\@. Likewise, we can consider the
contribution of the $\cA$-instanton by setting $B_1=0$. In this case
the transformation $\eta=\pi/4$, $\xi=\rho=0$ leads to a correction
 \begin{equation} \label{r.5}
  \Delta \tilde{\cW}_{1111} = 4 N A_1\, \e^{\I v}\ ,\qquad \Delta
  \tilde{\cW}_{2222} = 4 N A_1^*\, \e^{-\I v}\ ,
 \end{equation}
with the other components vanishing identically. This is again of the
form predicted by BBS, even though in a different fermionic frame
than \eqref{r.4}. Summing these two corrections involves rotating
some of the contributions into the proper fermionic frame. Hence, our
result precisely agrees with the one obtained in \cite{BBS}.

In order to sum the two contributions, we go to the $\cB$-instanton
frame, i.e. the frame in which we obtained \eqref{r.4}, but now 
we also include $A_1$. The corrections to the $\cW$-Tensor then are 
 \begin{align} \label{r.6}
  \Delta \tilde{\cW}_{1111} & = - N\, \big( A_1\, \e^{\I v} + A_1^*\,
    \e^{-\I v} + 4 B_1^*\, \e^{\I u} \big) \notag \\[2pt]
  \Delta \tilde{\cW}_{1112} & =  N\, \big( A_1\, \e^{\I v} - A_1^*\,
    \e^{-\I v} \big) \notag \\[2pt]
  \Delta \tilde{\cW}_{1122} & = - N\, \big( A_1\, \e^{\I v} + A_1^*\,
    \e^{-\I v} \big) \notag \\[2pt]
  \Delta \tilde{\cW}_{1222} & =  N\, \big( A_1\, \e^{\I v} - A_1^*\,
    \e^{-\I v} \big) \notag \\[2pt]
  \Delta \tilde{\cW}_{2222} & = - N\, \big( A_1\, \e^{\I v} + A_1^*\,
    \e^{-\I v} + 4 B_1\, \e^{-\I u} \big)\ .
 \end{align}

This result implies that the four fermionic zero modes $\psi_1$,
$\psi_2$ arising from a membrane wrapping the $\cA$ and the $\cB$ cycle,
respectively, are not orthogonal. If in the $\cB$-frame we denote the
two zero modes giving rise to the $\e^{\I u}$ corrections with $\psi_1$,
then the corrections proportional to $\e^{\I v}$ arise from the zero
modes $\psi_1-\psi_2$. This zero mode configuration produces all the
signs appearing in \eqref{r.6}, since the $\cA$-anti-instanton has its
zero modes in $\psi_1+\psi_2$.

\section{Constructing meta-stable de Sitter vacua} \label{dS}

We now move on and study the properties of the scalar potential that
arises from gauging the isometry of the Przanowski-Tod metric. We will
find that inclusion of the instanton corrections obtained in section
\ref{sect4} will lead to the stabilization of all hypermultiplet moduli
and also opens up the possibility for obtaining meta-stable dS vacua
from string theory. In order to highlight the role of the membrane
instanton corrections played in this construction, we will work with a
toy model of a rigid CY$_3$ compactification where we have truncated all
vector multiplets.\footnote{Based on the results obtained in this
section it is straightforward to adapt this model to any rigid CY$_3$
compactification by including the corresponding vector multiplet sector.
This also allows to study more general gaugings by turning on background
fluxes in the even cohomology classes of the CY$_3$.} The only remnant of
the vector multiplet sector will be that we allow for a non-trivial
(negative) one loop correction encoded by $c$.

\subsection{Turning on background fluxes} 

In $N=2$ supergravity the scalar potential arises from gauging
isometries of the scalar manifolds. In particular, the potential is
completely fixed once these manifolds are chosen and the gauged
isometries are specified.\footnote{This is different from $N=1$
supergravity where the potential depends on an arbitrary holomorphic
function, the superpotential.} For the instanton corrected UHM metric
the Heisenberg algebra of isometries present at the perturbative level
is broken explicitly and only the shift symmetry in $t$ remains. The
identification $t=\sigma$ (see discussion around \eqref{Heis-alg})
shows that gauging this isometry corresponds to gauging the shift
symmetry in the axion.

It is now natural to ask whether gauging this isometry has an
interpretation in terms of the 10-dimensional CY$_3$ compactification.
Indeed, comparing our gauging with the ones arising from flux
compactifications of the IIA string \cite{LM}, we find that it arises
from a non-trivial space-filling 3-form part $c_3$ of the RR 3-form
$C_3$. As pointed out in \cite{KKP}, this is equivalent to having
non-trivial 6-form flux $F_6$ in the CY$_3$\@. In four dimensions $c_3$
can be dualized to a constant $e_0$, which precisely leads to gauging
the shift in $t$; the Killing vector encoding the gauging is then given
by \eqref{k_t}. In the absence of vector multiplets gauging this
isometry induces the scalar potential (see appendix A for our
normalization and conventions)
 \begin{equation} \label{pot}
  V = 4\, G_{AB} k^A k^B - 3\, \vec{P}\! \cdot\! \vec{P} = \frac{1}
  {r^2}\, \big( 4 f^{-1} - 3 \big) e_0^2\ .
 \end{equation}
Here we have substituted the Killing vector \eqref{k_t} together
with the corresponding moment map \eqref{eq:moment1} and the PT metric
\eqref{1.1} in the second step.

Before discussing the properties of this potential, let us make some
remarks about the gauging. First, one might worry that the inclusion of
$F_6$ background flux could induce tadpoles, which would render our
model inconsistent. However, it was shown in \cite{KKP} that in a type
IIA compactification only turning on $F_0$ and $H_3$ flux simultaneously
gives rise to a tadpole condition, so that is not an issue here. Second,
as discussed in \cite{KKT} for the type IIB case, including background
fluxes can change the number of zero-modes that arise from a $p$-brane
instanton. But since $c_3$ has no support on the wrapped cycle, we do not
expect this to happen in our model, so that the membrane instantons will
still lead to a correction of the four-fermi coupling. Finally,
including background fluxes in a compactification in general leads to a
backreaction on the geometry of the internal manifold. These
backreactions are particularly relevant when looking for flux
compactifications which preserve (some) supersymmetry and, at the same
time, are consistent solutions of the 10-dimensional equations of
motion. In the case of CY$_3$ compactifications with non-trivial fluxes
this implies that the internal manifold should be a generalized CY$_3$
having an SU(3)-structure (see \cite{B} and references therein). We will
here neglect this backreaction of the flux on the geometry in the
following and tacitly assume that turning on fluxes will not drastically
alter the instanton results derived for a rigid CY$_3$ manifold in the
previous sections.

\subsection{The perturbative potential} 

\FIGURE{%
\epsfxsize=0.45\textwidth
\epsffile{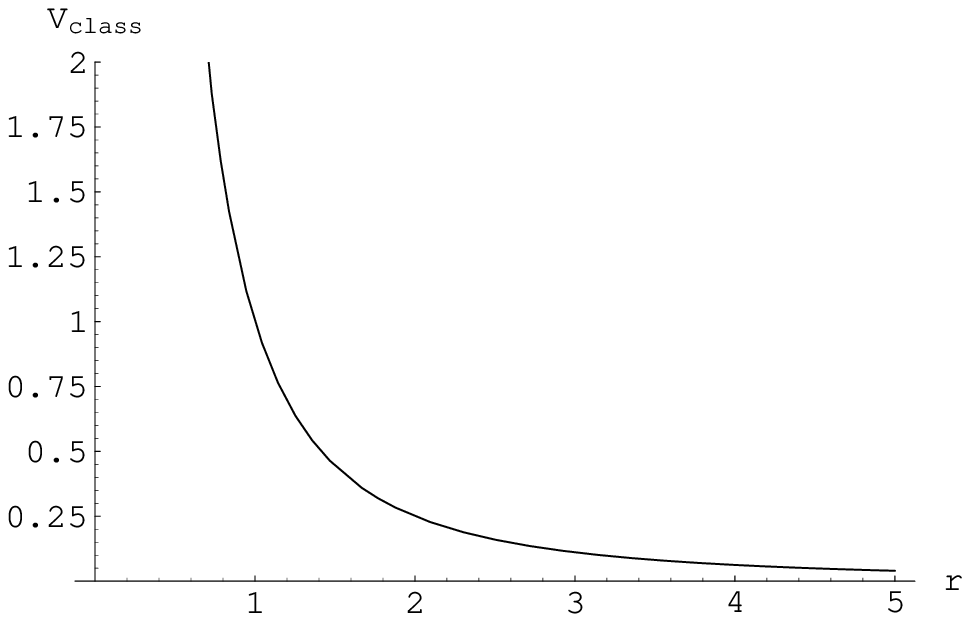}
\epsfxsize=0.45\textwidth
\epsffile{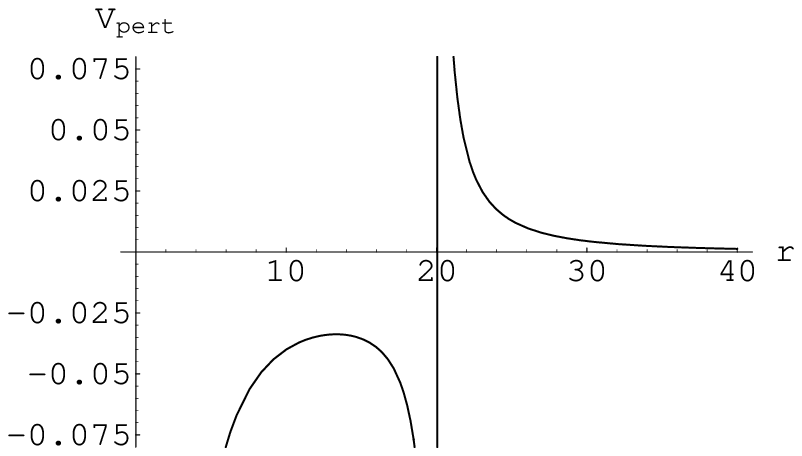}
\caption{\label{eins} The scalar potential $V_\class(r)$ (left) and
$V_\pert(r)$ (right) for $c=-10$. Including the perturbative corrections
with $c<0$ leads to a pole at $r=-2c$.}}

We now compute the scalar potential \eqref{pot} for the one-loop metric
\eqref{pertm}. Setting $e_0 = 1$ (which does not affect the vacuum
structure of the potential) we find
 \begin{equation} \label{pertpot}
  V_\pert = V_\class + V_\text{loop}\ ,
 \end{equation}
where
 \begin{equation}
  V_\text{class} =  \frac{1}{r^2}\ ,\qquad V_\text{loop} = -
  \frac{4c}{r^2 (r+2c)}\ .
 \end{equation}
Fig.\ \ref{eins} displays $V_\class$ and $V_\pert$, respectively, for
a ``typical'' value $c=-10$.

The classical potential shows a typical runaway behavior in $r$. It
is positive definite and diverges as $r\searrow 0$. For increasing $r$,
$V_\class$ decreases monotonically and there are no vacua except for the
trivial one at $r=\infty$ (vanishing string coupling). This is shown in
the left diagram of fig.~\ref{eins}.

Let us now add the $V_\text{loop}$-term to the scalar potential. The
sign of this contribution crucially depends on the sign of $c$, or
equivalently, on the Euler number of the Calabi-Yau.\footnote{For $c>0$,
$V_\pert$ goes to $-\infty$ as $r\searrow 0$. It then increases
monotonically up to $r=(1+\sqrt{5})c$, where it has an unstable
extremum, $V_\pert|_{r=(1+\sqrt{5})c} > 0$. For $r>(1+\sqrt{5})c$ the
potential decreases monotonically and approaches $V_\pert\searrow 0$
for $r\rightarrow\infty$.} The generic behavior of $V_\pert$ for $c<0$
is shown in the left diagram of fig.~\ref{eins}. Also in this case
$V_\pert=-\infty$ as $r\searrow 0$. In the interval $0<r<(1-\sqrt{5})c$
the potential increases monotonically. At $r=(1-\sqrt{5})c$ we again
find an unstable extremum $V_\pert|_{r=(1-\sqrt{5})c}=-(\sqrt{5}+1)/
(c^2(1+\sqrt{5})^2(3-\sqrt{5}))<0$. For $(1-\sqrt{5})c<r<-2c$ the
potential decreases monotonically and we obtain a second singularity at
$r=-2c$. For $r>-2c$, $V_\pert$ displays the runaway behavior already
found in the classical case. Notice, however, that in the region $0<r<
-2c$ the perturbatively corrected metric \eqref{1.1} is no longer
positive definite, so that this region does not belong to the moduli
space of the universal hypermultiplet.

It is important to note that the perturbative potential is independent
of $(u,v,t)$, so that these scalars correspond to flat directions. The
status of the flat directions corresponding to $(u,v)$ and $t$,
respectively, is quite different, however. This is due to the fact that
we have gauged the shift symmetry in $t$ (the axion). Gauge invariance
then requires that $t$ parameterizes a flat direction, which in turn
implies that one can gauge away the scalar $t$, giving a mass to the
gauge field, i.e., the vector field becomes massive by ``eating'' a
scalar via the St\"uckelberg mechanism. Therefore, only $u$ and $v$
have to be stabilized by the potential in order to fix all moduli. As
we will now show, this is readily achieved by including the leading
membrane instanton corrections in the scalar potential.

\subsection{The membrane-instanton contribution} 

We now demonstrate how the leading instanton correction can drastically
alter the vacuum structure of our low-energy effective action. To
illustrate this, let us consider the modifications arising from the
$v$-instanton sector only, while the (equally important) terms coming
from the $u$-instanton will be switched off for the sake of clarity.
Eq.\ \eqref{9.1} indicates that the contributions stemming from the one
$(u,v)$-instanton sector enter in exactly the same way. Hence, the
corrections arising from the one $u$-instanton can be included by taking
the $v$-dependent expressions given below, replacing $v$ by $u$ and
adding these additional terms to the potential. Therefore, it is clear
that our discussion for the $v$-modulus also applies to $u$. In
particular, the stabilization of the $v$-modulus can trivially be
extended to $u$ by including the $u$-instanton corrections as well.
Furthermore, the existence of a meta-stable dS vacuum is not limited to
the $u$-independent case and can also be obtained by including the
$u$-dependent terms in the potential. This will shift the boundaries for
the ``dS window'' discussed below to lower values of the integration
constants.

Let us now compute the leading contribution of a single $v$-instanton to
the scalar potential. 

Substituting $f$ given in \eqref{finst} into the potential \eqref{pot}, 
we find at leading order
 \begin{equation} \label{8.5}
  V_\text{1-inst} = -4\, r^{-(m_1+5)/2}\, \big( \hat{A}_{1,m_1} \cos(v)
  - \tilde{A}_{1,m_1} \sin(v) \big)\, \e^{-2 \sqrt{r}}\ .
 \end{equation}
Here we have set $A_{1,m_1}=\hat{A}_{1,m_1}+\I\tilde{A}_{1,m_1}$ and
$B_{1,m_1}=0$. Adding this contribution to the perturbative potential
\eqref{pertpot}, we then obtain in the semi-classical approximation
 \begin{equation} \label{8.7}
  V_\text{tot} = V_\text{class} + V_\text{loop} + V_\text{1-inst}\ .
 \end{equation}
The most important change arising from including $V_\text{1-inst}$ in
the potential is that the potential is no longer independent of $v$
(and, when including the $u$-instanton contribution, also of $u$).
Therefore the \emph{instanton correction lifts the $u,v$-degeneracy and
provides a non-perturbative mechanism to stabilize these moduli.}

Based on \eqref{8.7} we can make the following additional observations:
For $r\rightarrow\infty$ (vanishing $g_s$) all terms in $V_\text{tot}$
vanish, $\lim_{r\rightarrow\infty}V_\text{tot}=0$. For $r\gg 1$, 
$V_\text{tot}$ is dominated by its classical piece $V_\class\ge 0$, so
that $V_\text{tot}$ approaches zero from above. Furthermore,
$V_\text{1-inst}$ has no poles except at $r=0$.\footnote{In fact, this
is an artefact  of the expansion in \eqref{8.5}. If we do not expand the
denominator containing $(r+2c)$, $V_\text{1-inst}$ also develops a
singularity at $r=-2c$, which even dominates over the one in
$V_\text{loop}$, and the potential is no longer bounded from below.
Resolving this singularity presumably requires resumming the entire
instanton expansion to obtain expressions which are valid at small
values of $r\le -2c$. This resummation is, however, beyond the scope of
the present paper, and we will continue to work with the expanded
expressions \eqref{8.5}. Notice, however, that resolving singularities
by non-perturbative effects has been shown to work in the context of
the Coulomb branch of three-dimensional gauge theories with eight
supercharges \cite{SW,DKMTV}. In these cases the moduli space is
hyperk\"ahler instead of quaternion-K\"ahler.} Hence, for $r>0$
the only divergence in $V_\text{tot}$ is contained in $V_\text{loop}$,
which diverges at $r=-2c$. As a result, $V_\text{tot}$ is bounded from
below and diverges, $V_\text{tot}=+\infty$, as $r\searrow-2c$.

Analyzing the vacuum structure arising from $V_\text{tot}$ analytically
is rather difficult due to the transcendental nature of the potential.
We therefore have analyzed the vacuum structure using numerical methods.
Since we lack any better knowledge about the $g_s$-dependence of the
instanton measure plus 1-loop determinant around the single
$v$-instanton, we choose the lowest possible value $m_1=-2$. As it turns
out, the qualitative picture of the vacuum structure is not sensitive to
this choice.

In order to further simplify the potential \eqref{8.7}, we impose the
discrete $\fieldZ_2$ symmetry $v\rightarrow-v$, $t\rightarrow -t$
discussed in subsection \ref{sect4.3}. This symmetry can be made
manifest by setting $A_{1,m_1}=A^*_{1,m_1}$ or, equivalently,
$\tilde{A}_{1,m_1}=0$. Without loss of generality we can furthermore
choose $\hat{A}_{1,m_1}$ to be positive, as, in the leading order
approximation, considering negative values of $\hat{A}_{1,m_1}$ merely
corresponds to shifting $v\rightarrow v+\pi$. This choice of parameters
then implies that $v=0$ corresponds to a local minimum of the potential
in the $v$-direction.

\TABULAR[p]{c||c|c}{
  & $c = -1.9$ & $c = -10$ \\ \hline
  $A_\mn$ & $53.0$ & $8180$ \\
  $A_\mx$ & $60.8$ & $9900$ \\
  $r_\mathrm{dS}$ & $7.4$ & $27.5$}
{\label{t1} Illustrative values for
$A_\text{min}$, $A_\text{max}$, and $r_\mathrm{dS}$ for the
$\mathcal{Z}$-manifold ($h_{1,2}=0$, $h_{1,1}=36$, $c\approx-1.9$) and
a fictional rigid CY$_3$ where $c=-10$, corresponding to $h_{1,1}=
\cO(100)$. For $A_\mn<\hat{A}_{1,m_1}<A_\mx$ the potential \eqref{8.7}
has a meta-stable dS vacuum at which all hypermultiplet moduli are
stabilized.}

Depending on the value of the remaining free parameter $\hat{A}_{1,m_1}$
we obtain three classes of vacuum structures\footnote{Recall that we
consider rigid CY's, where $c<0$, and the region $r>-2c$ only.}, which
are separated by two ($c$-dependent) thresholds $\hat{A}_{1,m_1}=
A_\mn$ and $\hat{A}_{1,m_1}=A_\mx$. For $\hat{A}_{1,m_1}<A_\mn$ we find
the runaway behavior present in the classical and perturbative
potentials. In this case there is no vacuum, except for the trivial one
at $r=\infty$. For $A_\mx<\hat{A}_{1,m_1}$ on the other hand, we obtain
a stable AdS vacuum which is separated from the runaway vacuum by a
saddle point of the potential where $V_\text{tot}|_\text{saddle}>0$. In
this case all hypermultiplet moduli can be stabilized in the AdS vacuum.
The most interesting case, however, occurs for $A_\mn<\hat{A}_{1,m_1}<
A_\mx$. In this case the AdS vacuum is lifted to positive cosmological
constant and one obtains a \emph{meta-stable dS vacuum}. As in the AdS
case, this dS vacuum is separated from the runaway vacuum by a saddle
point of the potential where $V_\text{tot}|_\text{saddle}>0$.
\emph{This meta-stable dS vacuum stabilizes all the hypermultiplet
moduli}.

Furthermore, one can verify that increasing $\hat{A}_{1,m_1}$ results in
the (A)dS vacuum moving closer to the singularity at $r=-2c$. This
implies that the vacuum value (i.e., the value for which the string
coupling is weakest) of $r$ is obtained by setting $\hat{A}_{1,m_1}=
A_\mn$. In this case $r$ is stabilized in the meta-stable dS vacuum and
we will denote its corresponding value by $r_\mathrm{dS}$. Table
\ref{t1} then summarizes the values for $A_\mn$, $A_\mx$, and
$r_\mathrm{dS}$ for two ``typical'' values $c=-6/\pi\approx-1.9$ and
$c=-10$, respectively. The former corresponds to the
$\mathcal{Z}$-manifold (see e.g.\ \cite{CDP}), the prototype of a rigid
CY$_3$ with $h_{1,2}=0$ and $h_{1,1}=36$, while $c=-10$ reflects a
fictional rigid CY$_3$ where $h_{1,1}=\cO(100)$. Table \ref{t1}
indicates that decreasing $|c|$ also decreases the values for $A_\mn$
and $A_\mx$, while the relative width of the ``dS window'', $(A_\mx-
A_\mn)/(A_\mx+A_\mn)$, stays approximately constant. Furthermore, we
observe that decreasing $|c|$ moves $r_\mathrm{dS}$ closer to the
singularity at $r=-2c$.

\FIGURE{%
\epsfxsize=0.80\textwidth
\epsffile{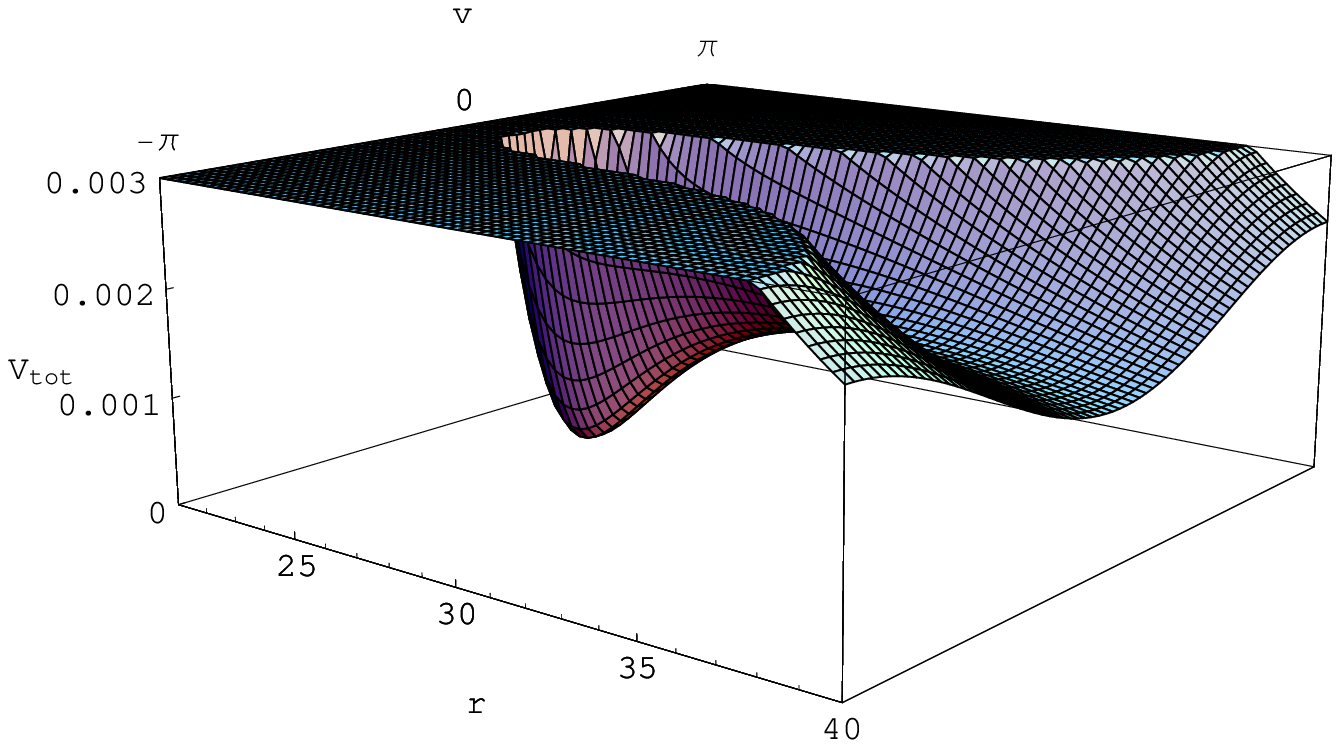}
\caption{\label{drei} A detailed view of the meta-stable dS minimum of
the potential $V_\text{tot}$.}}

Figs.\ \ref{drei} and \ref{zwei} show the typical shape of
$V_\text{tot}$ in the dS phase. Here we chose $c=-10$ and $\hat{A}_{1,
m_1}=9867$. Fig.\ \ref{drei} displays the $(r,v)$-dependence of the
potential, illustrating that we have indeed a meta-stable dS vacuum.
Fig.\ \ref{zwei} depicts $V_\text{tot}$ in the $r$-direction for $v=0$.

\FIGURE{%
\epsfxsize=0.60\textwidth
\epsffile{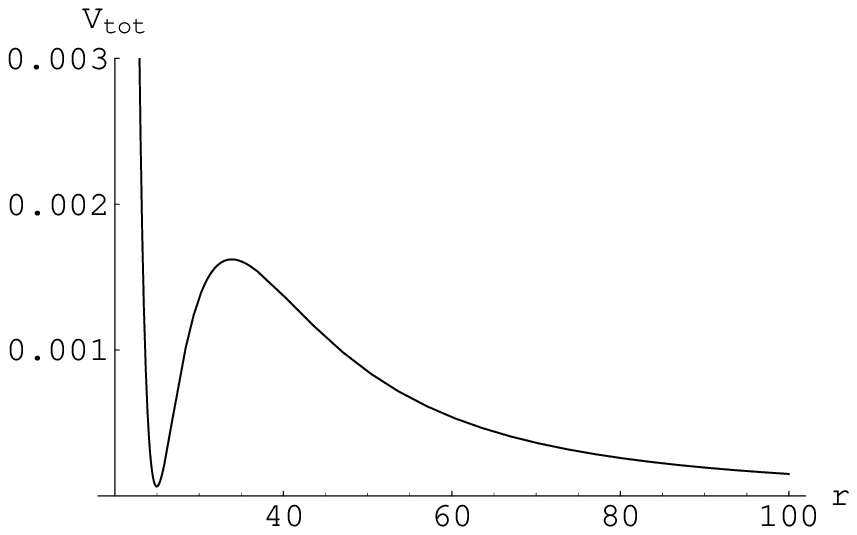}
\caption{\label{zwei} A view of $V_\text{tot}(r,v)$ in the $r$-direction
along $v=0$.}}

Let us conclude this section with a remark on the periodicity of
$V_\text{tot}$ in the $v$-direction, which arises from the oscillatory
terms in $V_\text{1-inst}$. As discussed in subsection \ref{sect4.3},
this reflects the fact that at leading order in the instanton correction
there is a residual $\fieldZ$ symmetry arising from the broken
$\beta$-shift. Strictly speaking, we then obtain an infinite number of
copies of the (A)dS vacua found above. Including higher order subleading
terms, however, will break this discrete symmetry completely, thereby
lifting the degeneracy between these vacua. In order to decide on the
fate of these vacua one would have to sum the whole instanton series,
which is, however, beyond the scope of this paper. We have verified
that, when performing the above analysis by taking into account the sub-
and subsubleading contributions to the potential \eqref{8.7}, one still
has (for a suitable choice of integration constants) one meta-stable dS
vacuum, while the other local minima generically become meta-stable AdS
vacua. This analysis gives some evidence that the qualitative picture
found above will remain valid after resumming the instanton series. In
particular, we expect that the stabilization of the hypermultiplets
will also be a feature of the complete instanton solution. 

To make further progress on this issue, we have to improve our
understanding on membrane instanton calculations beyond what has been
done in \cite{BBS,HM}. Ideally we would like to fix the numerical
coefficients in our instanton expansion completely through the
microscopic string theory description. For (some) more general CY$_3$
compactifications of the type IIA string this may be done using the
duality to heterotic string theory on $K3\times T^2$ \cite{FHSV}, where
the hypermultiplet moduli space is classically exact in the string
coupling constant. However, this duality generically involves more than
one hypermultiplet, which requires a more generic setup than what is
considered here. Furthermore, it would be interesting to investigate
whether these coefficients have some deeper meaning in the context of
topological string theory, analogous to the coefficients appearing in
the D3-brane instanton corrections to orientifold compactifications of
the type IIB string recently investigated in \cite{BM2}.

\section{Discussion}

Let us now discuss the relation between KKLT \cite{KKLT} and our set-up.
In order to stabilize all moduli and to obtain a meta-stable dS vacuum,
KKLT proposed a three step procedure, where first all moduli apart from
the dilaton were fixed by fluxes, second the dilaton is stabilized by
non-perturbative instanton effects at an AdS vacuum, and finally a
positive energy contribution (in form of anti-D3-branes) is added to
lift this vacuum to a meta-stable dS vacuum. When including a
space-filling RR 3-form flux in our case, the classical potential is
positive definite and of runaway type. There is no vacuum, except the
one at vanishing string coupling, and both RR scalars correspond to flat
directions. This does not change when the perturbative corrections to
the universal hypermultiplet found in \cite{AMTV} are included. The
picture changes completely when taking into account the leading membrane
instanton corrections to the universal hypermultiplet. These corrections
to the scalar metric lift the flat directions corresponding to the RR
scalars, so that all the present moduli are fixed by the potential.
Furthermore, by making a suitable choice of the numerical parameters
corresponding to the one-loop determinant around a one-instanton
background, the moduli can be stabilized in a meta-stable dS vacuum at
small string coupling $g_s\ll 1$. Here, the appearance of the dS vacuum
does not require to add a positive energy contribution (like
anti-D3-branes) by hand, as this contribution is already provided by
the background flux when taking the perturbative corrections to the
hypermultiplet geometry into account. This picture is completely
analogous to the one obtained for type the IIB orientifold
compactifications studied in \cite{BB2} where it was found that the
leading order $\alpha'$ corrections to the K\"ahler potential together
with the leading D3-brane instanton correction can also give rise to
meta-stable dS vacua without the need of adding an additional positive
energy contribution. On phenomenological grounds, it would be
interesting to analyze the stability and lifetime of these dS vacua
along the lines of \cite{KKLT}.

In our model we have truncated the vector multiplets that arise in a
realistic compactification of type IIA strings on a rigid Calabi-Yau
manifold. In order to address the issue of stabilizing these moduli as
well, it would be necessary to include them in the effective action.
This is, however, beyond the scope of the present paper. In this
context, let us remark that including these vector multiplets (whose
scalar fields correspond to the complexified K\"ahler moduli of the
compactification) also allows to consider more general fluxes, like
e.g.\ 2- and 4-form fluxes related to the even homology cycles of the
rigid Calabi-Yau manifold, which could then be used to stabilize these
moduli as well. Previous investigations on this topic indicate that
these moduli will likely be stabilized at special points of the K\"ahler
moduli space where the Calabi-Yau geometry degenerates (as e.g.\ at
conifold points), either through fluxes \cite{DL} or non-perturbative
effects arising from (Lorentzian) branes wrapping the degenerate cycles
\cite{CF}. In this context it is also interesting to note that instanton
corrections of the type discussed in this paper play an important role
in a proper understanding of string theory at these degeneration points
\cite{BBS,OV}. It would therefore be highly desirable to extend the
present analysis to include vector multiplets.

Possible extensions of the present work would be to include also
fivebrane instanton corrections to the universal hypermultiplet. Some
results were already obtained in \cite{DTV}, and it would be challenging
to combine them with the results obtained in this paper. Another
direction to pursue is to consider more generic Calabi-Yau threefolds,
in which one also has additional hypermultiplets. The
quaternion-K\"ahler manifold would then be higher dimensional, and it
would be interesting to study instanton effects in this context and see
how they stabilize all the complex structure moduli. We leave this open
for future research.
\bigskip

\acknowledgments{%
We thank Lilia Anguelova for collaboration in the early stages of the
project. We also thank Serguei Alexandrov, David Calderbank, Jan
de~Boer, Mathijs de~Vroome and Albrecht Klemm for stimulating
discussions. This project was initiated during the Simons Workshop
in Mathematics and Physics, Stony Brook 2004. UT was supported by the
DFG within the priority program SPP~1096 on string theory.}

\appendix

\section{Notation and conventions}

In order to study the scalar potential in gauged supergravity, we need
to fix the conventions and normalizations of the various quaternionic
quantities. We mainly follow the notation of \cite{DJDWKV,DWRV,DWRV2},
but with a few modifications on the conventions mentioned explicitly
below.

\subsection{General properties of quaternion-K\"ahler geometry} 
\label{appA.1}

The quaternionic structure is normalized such that\footnote{The
$J^\Lambda$ defined here differ from \cite{DWRV} by a minus sign.}
 \begin{equation} \label{2.1}
  J^\Lambda J^\Sigma = - \delta^{\Lambda\Sigma} - \varepsilon^{\Lambda
  \Sigma\Pi} J^\Pi\ ,
 \end{equation}
with $\Lambda,\Sigma,\Pi=1,2,3$. There exist quaternionic 1-form
vielbeine $V_i^\alpha$, in terms of which the line element reads
 \begin{equation} \label{2.2}
  \d s^2 = G_{AB}\, \d\phi^A \otimes \d\phi^B = G_{\bar{\alpha}\beta}
  V_i^\beta \otimes \bar{V}^{i\bar{\alpha}}\ .
 \end{equation}
Here, $G_{AB}$ is the quaternionic metric, $\bar{V}^{i\bar{\alpha}}$ is
the complex conjugate of $V_i^\alpha$, and $G_{\bar{\alpha}\beta}$ is
the tangent space metric that appears in front of the kinetic terms of
the fermions. The quaternionic 2-forms can then be written as
 \begin{equation}\label{2.3}
  J^\Lambda = \frac{\I}{2}\, G_{\bar\alpha \beta} V_i^\beta \wedge
  \bar{V}^{j\bar \alpha} (\tau^\Lambda)^i{}_j\ ,
 \end{equation}
where $\tau^\Lambda$ are the Pauli matrices. For $4n$-dimensional QK
manifolds, the range of the indices is $i=1,2$, $\alpha=1,...,2n$, and
$A=1,...,4n$.

Quaternion-K\"ahler manifolds are Einstein, and hence the Ricci tensor
is proportional to the metric. Following \cite{BCDGVV}, we have
 \begin{equation}
  R_{AB} = \frac{1}{4n}\, G_{AB} R\ ,
 \end{equation}
where $R$ is the (constant) Ricci scalar. Furthermore, there exist
SU(2) connection 1-forms $\vec{\cV}=\vec{\cV}_{\!A}\,\d\phi^A$ with
SU(2) curvature\footnote{The convention for the SU(2) connection and
curvature is chosen to be the same as e.g.\ in \cite{DWRV2}. With
respect to \cite{BCDGVV}, our SU(2) connection is chosen (minus) twice
the one in \cite{BCDGVV}, and therefore also the SU(2) curvature is
(minus) twice as large.}
 \begin{equation} \label{A.7}
  \vec{\cR} \equiv \d \vec{\cV} - \half\, \vec{\cV} \times \vec{\cV}\ .
 \end{equation}
The exterior derivative on $\vec{\cR}$ yields the Bianchi identities
 \begin{equation} \label{A.8}
  \d\vec{\cR} = \vec{\cV} \times \vec{\cR}\ .
 \end{equation}
The relation between SU(2) curvature and quaternionic 2-forms reads
 \begin{equation} \label{A.6}
  \vec{\cR} = \nu \vec{J}\ ,\qquad \nu \equiv \frac{1}{4n(n+2)}\, R\ .
 \end{equation}
For the gauging, we need the conventions for the moment maps. They are
defined from\footnote{Our definition of the moment map is the same as
in \cite{DWRV2}. This normalization is different from \cite{BCDGVV},
and our moment maps are (minus) two times the ones defined in
\cite{BCDGVV}.}
 \begin{equation} \label{def-mm}
  \vec{J}_{AB}\, k^B_I = D_A \vec{P}_I = (\p_A - \vec{\cV}_{\!A} \times)
  \vec{P}_I\ ,
 \end{equation}
where $I$ labels the different isometries and $D_A$ is the SU(2)
covariant derivative.

One can solve this relation for the moment maps to get
\cite{DWRV2} 
 \begin{equation} \label{2.9}
  \vec{P}_I = -\frac{1}{2n\nu}\, \vec{J}^A{}_B D_A k^B_I\ .
 \end{equation}
Notice that the right-hand side is independent of the metric, except
for the factor $\nu$. Choosing this factor sets the scale of the metric,
and for the universal hypermultiplet that we discuss below, we set the
scale\footnote{From the definition of $\nu$, it is clear that changing
the QK metric $G_{AB}\rightarrow\lambda G_{AB}$ changes the value of
$\nu$ according to $\nu\rightarrow\lambda^{-1}\nu$, while keeping the
first relation in \eqref{A.6} invariant.} such that $\nu=-1/2$.

In supergravity, the value of $\nu$ is fixed in terms of the
gravitational coupling constant. If we normalize the kinetic terms of
the graviton and scalars in the supergravity action as
 \begin{equation} \label{norm-action}
  e^{-1} \cL_\text{kin} = - \frac{1}{2\kappa^2}\, R(e) - \frac{1}{2}\,
  G_{AB} \p_\mu \phi^A\, \p^\mu \phi^B\ ,
 \end{equation}
then local supersymmetry fixes $\nu=-\kappa^2$. This is in accordance
with \cite{BCDGVV}, and with \cite{BW} after a rescaling of the metric
$G_{AB}$ with a factor 1/2. For the universal hypermultiplet, we will
work with conventions in which $\nu=-1/2$, so we set $\kappa^2=1/2$
below. To compare with \cite{DWRV2}, we first multiply the Lagrangian
\eqref{norm-action} by 2 and then set $\kappa^2=2$.

We now include the scalar potential that arises after gauging a single
isometry. The isometry can then be gauged by the graviphoton and in the
absence of any further vector multiplets, the relevant terms in the
Lagrangian are
 \begin{equation}
  e^{-1} \cL = - \frac{1}{2\kappa^2}\, R - \frac{1}{2}\, G_{AB} D_\mu
  \phi^A\, D^\mu \phi^B - \big( 2 \kappa^{-2} G_{AB}\, k^A k^B - 3\,
  \vec{P}\! \cdot\! \vec{P}\, \big)\ .
 \end{equation}
Here, $D_\mu$ is the covariant derivative with respect to the gauged
isometry that corresponds to the Killing vector $k^A$. The factors of
$\kappa$ appear on dimensional grounds, as one can easily verify. For
$\kappa^2=2$ this agrees precisely with the result in \cite{DWRV2};
here, however, we set $\kappa^2=1/2$.

Our conventions are chosen such that they naturally apply to the
universal hypermultiplet metric and the conventions used in \cite{TV2}.
At the classical level we have
 \begin{equation}
  \d s^2 = G_{AB}\, \d\phi^A \otimes \d\phi^B = \d\phi^2 + \e^{-\phi}
  (\d\chi^2 + \d\varphi^2) + \e^{-2\phi} (\d\sigma + \chi \d\varphi
  )^2\ .
 \end{equation}
For the corresponding Ricci tensor we find
 \begin{equation}
  R_{AB}= - \frac{3}{2}\, G_{AB}\ .
 \end{equation}
The Ricci scalar is then $R=-6$ and therefore we have $\nu=-1/2$. This
implies that in these conventions we should set $\kappa^2=1/2$, which
is equivalent to a cosmological constant $\Lambda=-3/2$ on the 
quaternion-K\"ahler manifold.

\subsection{Quaternion-K\"ahler geometry of the PT metric} 
\label{appA.2}

The quaternionic properties of the PT metric can be demonstrated by
constructing the corresponding quaternionic 1-form vielbeine
\eqref{2.2}, which we parameterize as
 \begin{equation}
  V_i^\alpha = \lp \bar{a} & - \bar{b} \\ b & a \rp\ .
 \end{equation} 
Substituting this ansatz into \eqref{2.2}, we obtain
 \begin{equation} \label{eq:metric1}
  \d s^2 = a \otimes \bar{a} + b \otimes \bar{b} + \text{c.c.}\ .
 \end{equation}
Comparing this expression with the PT metric \eqref{1.1}, we can choose
 \begin{equation}
  a = \frac{1}{\2\, r}\, \big( f^{1/2}\, \d r + \I f^{-1/2}\, (\d t +
  \Theta) \big)\ ,\qquad b = \frac{1}{\2\, r}\, (f \e^h)^{1/2}\,
  \big( \d u + \I\, \d v \big)\ .
 \end{equation}  
The computation of the quaternionic 2-forms \eqref{2.3} then yields
 \begin{equation}
  J^1 = -\I (a \wedge b - \bar{a} \wedge \bar{b})\ ,\quad
  J^2 = a \wedge b + \bar{a} \wedge \bar{b}\ ,\quad
  J^3 = -\I (a \wedge \bar{a} + b \wedge \bar{b})\ .
 \end{equation}
These satisfy the quaternionic algebra \eqref{2.1}.

Using \eqref{A.6} and \eqref{A.8}, we then determine the SU(2)
connection for the PT metric,
 \begin{gather}
  \mathcal{V}^1 = \frac{1}{r}\, \e^{h/2} \d v\ ,\qquad \mathcal{V}^2
    = \frac{1}{r}\, \e^{h/2} \d u\ , \notag \\
  \mathcal{V}^3 = -\frac{1}{2r}\, (\d t + \Theta) - \frac{1}{2}\,
    (\p_v h\, \d u - \p_u h\, \d v)\ .
 \end{gather}

The PT metric has a shift symmetry in $t$. In coordinates $(r,u,v,t)$
the corresponding Killing vector is given by
 \begin{equation}
  k^A = (\, 0\, ,\, 0\, ,\, 0\, ,\, e_0\,)^\mathrm{T}\ .
 \end{equation} 
The moment maps of this shift symmetry can be computed from \eqref{2.9}.
The result is \emph{independent} of the functions $f$, $h$, and $\Theta$
and reads
 \begin{equation}
  P^1 = 0\ ,\qquad P^2 = 0\ ,\qquad P^3 = \frac{e_0}{r}\ .
 \end{equation}

\subsection{The 4-fermion coupling of the PT metric} 
\label{appA.3}

In order to make contact with the string calculation of \cite{BBS}, we
need to construct the symmetric tensor $\cW_{\alpha\beta\gamma\delta}$,
which appears in the four-fermion term. This can be done along the lines
outlined in \cite{BCDGVV}. Note that the tensor
$\Omega_{XYZW}$ appearing in Bagger and Witten \cite{BW} is totally
symmetric for rigid supersymmetry, but \emph{not} in the supergravity
case.

The symmetric tensor $\cW_{\alpha\beta\gamma\delta}$ can be obtained
from the curvature decomposition  
 \begin{equation} \label{r.1}
  R_{ABCD} = \nu (R^{\SU(2)})_{ABCD} + \frac{1}{2}\, {L_{DC}}^{\alpha
  \beta}\, \cW_{\alpha\beta\gamma\delta}\, {L_{AB}}^{\gamma\delta}\ .
 \end{equation}
Here,
 \begin{equation}
  (R^{\SU(2)})_{ABCD} = \frac{1}{2}\, g_{D[A}\, g_{B]C} + \frac{1}{2}\,
  J^\Lambda_{AB}\, J^\Lambda_{DC} - \frac{1}{2}\, J^\Lambda_{D[A}\,
  J_{B]C}^\Lambda\ ,
 \end{equation}
and
 \begin{equation}
  {L_{AB\alpha}}^\beta = V_{A i\alpha}\, \bar{V}^{i\beta}_B\ .
 \end{equation}
Eq.\ \eqref{r.1} can be solved for $\cW_{\alpha\beta\gamma\delta}$ by
using the inverse relation for ${L_{AB}}^{\alpha\beta}$:
 \begin{equation}
  - \frac{1}{2}\, V^{iB}_\gamma\, V^{A}_{i\delta}\, {L_{AB}}^{\alpha
  \beta} = \delta^\alpha_\gamma\, \delta^\beta_\delta\ .
 \end{equation}
The resulting expression for $\cW_{\alpha\beta\gamma\delta}$ then reads:
 \begin{equation}
  \cW_{\alpha\beta\gamma\delta} = \frac{1}{2}\, \epsilon^{ij}
  \epsilon^{kl}\, V_{i\delta}^A\, V_{j\gamma}^B\, V_{k\beta}^D\, V_{l
  \alpha}^C \big( R_{ABCD} - \nu (R^{\SU(2)})_{ABCD} \big)\ .
 \end{equation}
The components of $\cW_{\alpha\beta\gamma\delta}$ can now be
obtained by calculating $R_{ABCD}$ for the PT metric \eqref{1.1} and
substituting the expressions for the vielbeins and complex structures
obtained above in the corresponding definitions. In order to write the
independent components of $\cW_{\alpha\beta\gamma\delta}$ in a compact
way, it is useful to introduce the complex variable $z=u+\I v$. The
result is given in \eqref{r.2}.

\section{Tensor multiplet description}

Consider the hypermultiplet Lagrangian based on the PT metric
\eqref{1.1}. It is interesting to write down the $N=2$ tensor multiplet
Lagrangian obtained after dualizing the scalar $t$ into a 2-form gauge
potential with field strength $H_{\mu\nu\rho}$. Using the results of
\cite{TV2}, this Lagrangian can easily be read off,
 \begin{equation}
  \cL_T = \frac{1}{2}\, r^2 f H_\mu H^\mu - \frac{1}{2}\, \cG_{AB}\,
  \p_\mu \phi^A\, \p^\mu \phi^B - \Theta_A H^\mu \p_\mu \phi^A\ .
 \end{equation}
Here, $\Theta_A$ are the three components of the one-form defined in
\eqref{dT}, and $\cG_{AB}$ is the metric on the manifold spanned by the
three scalars $(r,u,v)$. The line element can be written as
 \begin{equation} \label{LB-metric}
  \d s^2 = \frac{f}{r^2}\, \Big[ \d r^2 +  \e^h \big( \d u^2 + \d v^2
  \big) \Big]\ ,
 \end{equation}
where $\e^h$ satisfies the Toda equation and $f(r,u,v)$ the constraint
\eqref{f-h}. This 3-dimensional geometry is related to Einstein-Weyl
spaces, as explained in \cite{W}.

\section{Details of the Toda solution} \label{AppC}

This appendix collects several technical details about the solution of
the Toda equation constructed in section \ref{sect4}. We start by
proving $m_n\ge -2$ in subsection \ref{AppC1}, while the proof for
$\alpha=0$ is given in subsection \ref{AppC2}. The derivation of the
one-instanton solution is given in subsection \ref{AppC3}.

\subsection{The lower bound on $m_n$} 
\label{AppC1}

In this subsection we establish $m_n\ge -2$. Our starting point is the
ansatz \eqref{Ansatz}, which we substitute into the Toda equation
\eqref{Toda2}. This results in the following power series
expansion\footnote{Here we have not performed the splitting into
instanton sectors yet.}
 \begin{align} \label{C.10}
  0 = & \sum_{n,m} r^{-m/2+\alpha+1}\, \e^{-2n\sqrt{r}} \big[ (\Delta +
	n^2) f_{n,m} + (n\, a_{m+1}\, r^{-1/2} + b_{m+2}\, r^{-1})
	f_{n,m} \big] \notag \\[2pt]
  & + \sum_{n,m} \sum_{n',m'} r^{-(m+m')/2 + 2\alpha}\, \e^{-2(n+n')
	\sqrt{r}}\, \big[ f_{n',m'} (\Delta + 2n^2) f_{n,m}  \notag
	\\[-6pt]
  & \mspace{128mu} - \nabla f_{n,m} \cdot \nabla f_{n',m'} + 2 (a_{m+1}
	\, r^{-1/2} + b_{m+2}\, r^{-1}) f_{n,m}\, f_{n',m'} \big] \notag
	\\[8pt]
  & + \sum_{n,m} \sum_{n',m'} \sum_{n'',m''} r^{-(m+m'+m'')/2 + 3\alpha
	-1}\, \e^{-2 (n+n'+n'') \sqrt{r}} f_{n,m}\, f_{n',m'}\, f_{n'',
	m''} \notag \\[-8pt]
  & \mspace{128mu} \times \big[ n^2 + n\, a_{m+1}\, r^{-1/2} + b_{m+2}\,
	r^{-1} \big]\ ,
 \end{align}
where we have extended the definitions for $a_m$, $b_m$ given in
\eqref{4.6} to non-zero $\alpha$:
 \begin{equation} \label{C.2}
  a_m = \half\, (2m - 4\alpha - 1)\ ,\qquad b_m = \quart\, (m - 2\alpha)
  (m - 2\alpha - 2)\ .
 \end{equation}

In order to obtain a bound on $m_n$ (for which the $f_{n,m_n}\neq 0$),
we extract the leading order contributions in the $r$-expansion arising
from the single, double and triple sum in \eqref{C.10}. Starting at
$n=1$ and working iteratively towards higher values $n=2,3,\ldots$,
we find that at a fixed value of $n$ these contributions are
proportional to
 \begin{align} \label{C.11}
  \text{single sum}\ & \propto\ r^{-m_n/2 + \alpha + 1} \notag \\[2pt]
  \text{double sum}\ & \propto\ r^{-m_n + 2 \alpha} \notag \\[2pt]
  \text{triple sum}\ & \propto\ r^{-3 m_n /2 + 3 \alpha - 1}\ .
 \end{align}
Investigating the $m_n$-dependence of these relations, we find that for
$m_n\le -3$ the leading order term in $r$ arises from the triple sum,
which decouples from all the other terms in \eqref{C.10}.

We now assume that for a fixed value $n$ there exsists an $f_{n,m_n}\neq
0$ for $m_n\le -3$. Extracting the equation leading in $r$ from
\eqref{C.10}, we find that
 \begin{equation} \label{C.12}
  n^2\, f^3_{n,m_n} = 0\ ,\qquad m_n \le - 3\ ,
 \end{equation}
which has $f_{n,m_n}=0$ as its only solution. Hence, we establish the
lower bound
 \begin{equation} \label{C.13}
  m_n \ge -2
 \end{equation}
for all values of $n$ or, equivalently, all instanton
sectors.\footnote{Notice that this argument is not quite sufficient to
also fix $\alpha=0$, as for $\alpha=1/4$ the single and triple sums do
not decouple, which has been crucial in establishing \eqref{C.12}.}

\subsection{Fixing the parameter $\alpha$} 
\label{AppC2}

When making the ansatz \eqref{Ansatz} in order to describe membrane
instanton corrections to the universal hypermultiplet, we included the
parameter $\alpha\in[0,1/2[$ to allow for the possibility that the
leading term in the instanton solution occurs with a fractional power of
$g_s$. Based on the plausible assumption that the perturbation series
around the instanton gives rise to a power series in $g_s$ (and not
fractional powers thereof) we now give a proof that a consistent
solution of the Toda equation requires $\alpha=0$.

Splitting \eqref{C.10} into instanton sectors gives us the following
analogue of \eqref{4.5}
 \begin{align}\label{C.1}
  0 = \sum_{n,m} &\ r^{-m/2+\alpha}\, \e^{-2n\sqrt{r}}\, \Big\{ (\Delta
	+ n^2)\, f_{n,m+2} + n\, a_{m+2}\, f_{n,m+1} + b_{m+2}\, f_{n,m}
	\notag \\[-2pt]
  & + \sum_{n',m'} r^{\alpha}\, \e^{-2n'\!\sqrt{r}}\, \big[\, 2n\,
	a_{m'+1}\, f_{n',m-m'-1} + 2 b_{m'+2}\, f_{n',m-m'-2} \notag
	\\[-6pt]
  & \mspace{134mu} + f_{n',m-m'}\, (\Delta + 2n^2) - \nabla f_{n',m-m'}
	\cdot \nabla \big] f_{n,m'} \notag \\[4pt]
  & + \sum_{n',m'} \sum_{n'',m''} r^{2\alpha}\, \e^{-2(n'+n'')
	\sqrt{r}}\, f_{n,m'} f_{n',m''}\, \big[ n^2 f_{n'',m-m'-m''-2}
	\notag \\[-8pt]
  & \mspace{94mu} +  n\, a_{m'+1}\, f_{n'',m-m'-m''-3} + b_{m'+2}\,
	f_{n'',m-m'-m''-4} \big] \Big\} \, .
 \end{align}

Based on this equation we can now make several observations. First, we
find that the $N=1$ sector of \eqref{C.1} still gives rise to
\eqref{f1m}, with the coefficients $a_m$, $b_m$ now replaced by
\eqref{C.2}. To lowest order, $m=m_1$, this is just the equation
 \begin{equation} \label{C.3}
  (\Delta + 1) f_{1,m_1}(u,v) = 0\ .
 \end{equation}
Second, we observe that the equation describing the $N=2$ sector is
modified to
 \begin{align*}
  0 & = (\Delta + 4)\, f_{2,m} + 2a_m\, f_{2,m-1} + b_m\, f_{2,m-2} \\
  & \tab + \sum_{m'} r^{\alpha} \big[ f_{1,m-m'-2} + a_{m'+1}\,
	f_{1,m-m'-3} + b_{m'+2}\, f_{1,m-m'-4} - \nabla f_{1,m-m'-2}
	\cdot \nabla \big] f_{1,m'}\ .
 \end{align*}
Note that for $\alpha=0$ the sum appearing in the second line is just
an inhomogeneous term to the equations determining $f_{2,m}$. For $\alpha
\neq 0$, however, the sum decouples due to the different powers in $r$.
Therefore, in the case $\alpha\neq 0$, the sum gives rise to an
additional constraint equation, which is absent for $\alpha=0$. Since
the sum contains the $f_{1,m}$ only, this additional relation imposes a
restriction on the $N=1$ instanton solution. Upon using \eqref{C.3},
this additional constraint reads, at the lowest level,
 \begin{equation} \label{C.5}
  f_{1,m_1}^2 - (\nabla f_{1,m_1})^2 = 0\ .
 \end{equation}
For $\alpha\neq 0$ a non-trivial 1-instanton solution has to satisfy
both \eqref{C.3} and \eqref{C.5}, so that for establishing $\alpha=0$
it suffices to show that these equations have no common non-trivial
solution:

Suppose that $f_{1,m_1}\neq 0$, which by definition of $f_{1,m_1}$ has
to hold. We then multiply \eqref{C.3} with $f_{1,m_1}$, giving
 \begin{equation*}
  0 = f_{1,m_1}\, \Delta f_{1,m_1} + f_{1,m_1}^2 = f_{1,m_1}\, \Delta
  f_{1,m_1} + (\nabla f_{1,m_1})^2 = \half \Delta f^2_{1,m_1}\ ,
 \end{equation*}
where we have used \eqref{C.5} in the first step. In terms of complex
coordinates $z=u+\I v$ it is $\Delta=4\p_z\p_{\bar{z}}$, and the general
solution reads
 \begin{equation*}
  f_{1,m_1}^2(z,\bar{z}) = g(z) + \bar{g}(\bar{z})\ .
 \end{equation*}
Substituting this back into \eqref{C.3}, we find
 \begin{equation*}
  0 = (\Delta + 1) f_{1,m_1} = f_{1,m_1}^{-3} \big[\! - \partial_z
  g(z)\, \p_{\bar{z}} \bar{g}(\bar{z}) + (g(z) + \bar{g}(\bar{z}))^2
  \big]\ ,
 \end{equation*}
which is equivalent to
 \begin{equation*}
  \p_z g(z)\, \p_{\bar{z}} \bar{g}(\bar{z}) = g(z)^2 + 2 g(z) \bar{g}
  (\bar{z}) + \bar{g}(\bar{z})^2\ .
 \end{equation*}
Since the right-hand side of this expression contains terms which are
(anti-) holomorphic, whereas the left-hand side does not, we find that
the only solution is given by $g(z)=\I c$ with $c\in\fieldR$ constant.
Thus $f_{1,m_1}=0$, which contradicts our assumption and shows that the
ansatz \eqref{Ansatz} does \emph{not} give rise to a one-instanton
sector if $\alpha\neq 0$. Conversely, a non-trivial one-instanton sector
exists for $\alpha=0$ only, which then fixes $\alpha=0$.

\subsection{The one-instanton solution} 
\label{AppC3}

The general one-dimensional solution in the one-instanton sector was
given in \eqref{solf1m}. The functions $G_s(x)$ introduced there are
defined by
 \begin{equation}
  G_s(x) = x^{s+1} h_{s-1}(x)\ ,
 \end{equation}
where $h_s(x)=j_s(x)+\I y_s(x)$ are the spherical Bessel functions of
the third kind. For $s\geq 0$ the $G_s(x)$ have no poles. Explicitly,
they read
 \begin{equation}
  G_0(x) = \e^{\I x}\ , \qquad G_{s>0}(x) = 2^{-s}\, \e^{\I x}\,
  \sum_{k=1}^{s} \frac{(2s-k-1)!}{(s-k)!\, (k-1)!}\, (-2\I x)^k\ .
 \end{equation}
Using the properties
 \begin{equation}
  x^2 h_s'' + 2x\, h_s' + \big[ x^2 - s(s+1) \big] h_s = 0\ ,\qquad
  h_s' + \frac{s+1}{x}\, h_s = h_{s-1}\ ,
 \end{equation}
we easily verify the relation
 \begin{equation}
  (\p_x^2 + 1)\, G_s(x) = 2s\, G_{s-1}(x)\ .
 \end{equation}
The proof of \eqref{solf1m} is now simple:
 \begin{align}
  (\p_x^2 + 1) f_{1,m}(x) & = \re \sum_{s\geq 0}\, \frac{1}{s!\,
	(-2)^s}\, k_{1,m}(s)\, (\p_x^2 + 1) G_s(x) \notag \\
  & = - \re \sum_{s\geq 1}\, \frac{1}{(s-1)!\,(-2)^{s-1}}\, k_{1,m}
	(s)\, G_{s-1}(x) \notag \\
  & = - \re \sum_{s\geq 0}\, \frac{1}{s!\,(-2)^s}\, k_{1,m}(s+1)\,
	G_s(x) \notag \\
  & = - \re \sum_{s\geq 0}\, \frac{1}{s!\,(-2)^s}\, \big[ a_m
	k_{1,m-1}(s) + b_m k_{1,m-2}(s) \big]\, G_s(x) \notag \\ 
  & = -\, a_m f_{1,m-1}(x) - b_m f_{1,m-2}(x)\ . \label{proof1}
 \end{align}
For the general $(u,v)$-dependent solution given in \eqref{solf2m}, the
proof is almost identical.

\raggedright 

\end{document}